%% file: arxiv.tex
\documentclass[]{jfm}

\usepackage{graphicx}
\usepackage{newtxtext}
\usepackage{newtxmath}
\usepackage{natbib}
\usepackage{hyperref}
\hypersetup{
    colorlinks = true,
    urlcolor   = blue,
    citecolor  = blue,
}

\newcommand{\RomanNumeralCaps}[1]
\linenumbers

\graphicspath{{FIGURES/}}
\usepackage{amsmath}
\usepackage{calrsfs}
\DeclareMathOperator{\pe}{\mathit{Pe}}
\DeclareMathOperator{\pr}{\mathit{Pr}}
\DeclareMathOperator{\sch}{\mathit{Sc}}
\DeclareMathOperator{\Sh}{\mathit{Sh}}
\DeclareMathOperator{\nus}{\mathit{Nu}}

\DeclareMathOperator{\ra}{\mathit{Ra}}
\DeclareMathOperator{\da}{\mathit{Da}}
\newcommand*\diff{\mathop{}\!\mathrm{d}}

\newcommand{\bu}{\boldsymbol{u}}
\newcommand{\grad}{\boldsymbol{\nabla}}

\usepackage[justification=justified, width=\textwidth]{caption}
\usepackage{multirow}
\usepackage{booktabs}
\usepackage{afterpage}
\usepackage{color}
\definecolor{ForestGreen}{RGB}{34,139,34}

\usepackage{mathrsfs}
\usepackage{cancel}
\usepackage{layouts}

\title{Towards the understanding of convective dissolution in confined porous media: thin bead pack experiments, two-dimensional direct numerical simulations and physical models}

\author{Marco De Paoli\aff{1,2}\corresp{\email{m.depaoli@utwente.nl}}, 
Christopher J. Howland\aff{1}\corresp{\email{c.j.howland@utwente.nl}}, 
Roberto Verzicco\aff{1,3,4}\corresp{\email{verzicco@uniroma2.it}} \and 
Detlef Lohse\aff{1,5} \corresp{\email{d.lohse@utwente.nl}}}
\affiliation{
\aff{1}Physics of Fluids Group and Max Planck Center for Complex Fluid Dynamics and J. M. Burgers Centre for Fluid Dynamics, University of Twente, P.O. Box 217, 7500AE Enschede, The Netherlands
\aff{2}Institute of Fluid Mechanics and Heat Transfer, TU Wien, 1060 Vienna, Austria
\aff{3}Dipartimento di Ingegneria Industriale, University of Rome ``Tor Vergata'', 00133 Roma, Italy
\aff{4}Gran Sasso Science Institute, 67100 L'Aquila, Italy
\aff{5}Max Planck Institute for Dynamics and Self-Organization, 37077 Gottingen, Germany
}

\begin{document}
\maketitle

\begin{abstract}
We consider the process of convective dissolution in a homogeneous and isotropic porous medium. The flow is unstable due to the presence of a solute that induces a density difference responsible for driving the flow. The mixing dynamics is thus driven by a Rayleigh-Taylor instability at the pore scale. We investigate the flow at the scale of the pores using Hele-Shaw type experiment with bead packs, two-dimensional direct numerical simulations and physical models. Experiments and simulations have been specifically designed to mimic the same flow conditions, namely matching porosities, high Schmidt numbers, and linear dependency of fluid density with solute concentration. In addition, the solid obstacles of the medium are impermeable to fluid and solute. We characterise the evolution of the flow via the mixing length, which quantifies the extension of the mixing region and grows linearly in time. The flow structure, analysed via the centre-line mean wavelength, is observed to grow in agreement with theoretical predictions. Finally, we analyse the dissolution dynamics of the system, quantified through the mean scalar dissipation, and three mixing regimes are observed. Initially, the evolution is controlled by diffusion, which produces solute mixing across the initial horizontal interface. Then, when the interfacial diffusive layer is sufficiently thick, it becomes unstable, forming finger-like structures and driving the system into a convection-dominated phase. Finally, when the fingers have grown sufficiently to touch the horizontal boundaries of the domain, the mixing reduces dramatically due to the absence of fresh unmixed fluid. With the aid of simple physical models, we explain the physics of the results obtained numerically and experimentally. The solute evolution presents a self-similar behaviour, and it is controlled by different length scales in each stage of the mixing process, namely the length scale of diffusion, the pore size, and the domain height.
\end{abstract}


\section{Introduction}\label{sec:rt_intro}

Natural convective flows take place when density differences within a fluid domain drive the fluid motion. 
Natural convection is the dominant heat and mass transport mechanism for many flows of practical interest, both in nature \citep{hartmann2001tropical,marshall1999open,cardin1994chaotic} and in industrial applications \citep{bejan2013convection,krepper2002investigation}.
One of the particularly relevant ones in geophysics is the case of convection in porous media \citep{depaoli2023convective}, in which the fluid occupies the interstitial space of a solid porous matrix. 
This configuration is typical of subsurface transport phenomena, and is relevant for petroleum migration \citep{simmons2001variable}, water contamination \citep{leblanc1984sewage,molen1988}, underground hydrogen storage \citep{zivar2021underground,krevor2023subsurface}, superficial formations in dry salty lakes \citep{lasser2021stability,lasser2023salt}, and sea ice growth \citep{wettlaufer1997,feltham2006,middleton2016}, to name a few.

Recently, this problem has been subject of extended investigations due to the implications it can bear for geological carbon sequestration \citep{huppert2014fluid,emami2015convective}.
In this process, carbon dioxide (CO$_{2}$) is injected into underground formations located one to three kilometres beneath the earth surface, with the aim of permanent storage.
After injection, carbon dioxide remains in a supercritical state due to the high pressure at the injections depths, but its density is lower compared to that of the resident fluid (brine), and the injected volume of carbon dioxide migrates on top of the brine layer \citep{macminn2012spreading}.
This situation is undesired, because it may cause a further migration of carbon dioxide to the upper layers, eventually leading to a return of CO$_{2}$ into the atmosphere.
However, carbon dioxide is partially soluble in brine, and when these fluids mix a heavier solution (CO$_{2}$ + brine) forms.
When a sufficiently thick layer of this denser mixture builds up, it may become unstable and form convective instabilities \citep{depaoli2021influence}.
These flow structures have a favourable effect on the efficiency of the CO$_{2}$ dissolution mechanism, since they contribute to fluid mixing (i.e., CO$_{2}$ permanently stored in brine) in a much more effective manner compared to pure diffusion \citep{ennis2005onset,xu2006convective}.
On the other hand, the process is made complex by the presence of these non-linear instabilities, and making reliable predictions of the dissolution dynamics is a non-trivial task.

The dissolution of CO$_{2}$ in brine is a problem characterised by an integral length scale corresponding to the height of the reservoir, which may be of the order of tens or hundreds of meters in geological formations of practical interest \citep{huppert2014fluid}. 
Therefore, resolving all the flow scales, from the pores to the reservoir scale, is not an option with the present computational and experimental capabilities.
A possible approach commonly employed to revolve the large scales of the flow consists of analysing the flow at the Darcy scale, i.e. the flow is not resolved within the pores, but the quantities of interest (pressure, velocity, concentration, temperature) are obtained as averaged over a representative volume containing many pores \citep{nield17}.
The majority of the works on convective flows in porous media currently available in literature refers to this case.

The mixing process of CO$_{2}$ and brine is idealised considering an homogeneous porous slab with a constant concentration difference applied at the horizontal boundaries (Rayleigh-B\'enard), or as a semi-infinite (one-sided) domain where the concentration is fixed at top and no flux is applied at bottom boundary \citep{hewitt2020vigorous}.
In the Rayleigh-B\'enard-Darcy case, the system attains a statistically steady-state, which has been predicted theoretically \citep{horton1945convection,lapwood1948convection} and recently accurately described numerically in two \citep{hewitt2012ultimate,depaoli2016influence,wen_2018,ulloa2022energetics} and three dimensions \citep{hewitt2014high,depaoli2022strong,dhar2022convective}, also at high Rayleigh-Darcy numbers, $\ra^*$ (the Rayleigh-Darcy number indicates the relative importance of convective and diffusive contributions in a flow through a porous medium).
The one-sided-Darcy configuration is characterised by a transient behaviour \citep{riaz2006onset,fu2013pattern} consisting of three main phases: (i) an initial diffusive regime in which the mixing layer grows and becomes eventually unstable \citep{depaoli2017solute}, (ii) a convective phase characterised by a constant dissolution rate \citep{hidalgo2012scaling}, and (iii) a finite-size regime when the strength of convection reduces gradually
\citep{slim2013dissolution}.

The Rayleigh-B\'enard-Darcy and the one-sided-Darcy systems have been well characterised numerically and theoretically in case of thermal convection at the Darcy scale, i.e. when the flow structures are large compared to the scale of the pores \citep{hewitt2020vigorous}.
Numerical results suggest that a linear relationship exists between the dimensionless mass transfer coefficient (Sherwood number, $\Sh$) and the Rayleigh-Darcy number ($\ra^*$), in both the Rayleigh-B\'enard-Darcy case \citep{hewitt2012ultimate,pirozzoli2021towards} and during the convective regime of the one-sided-Darcy system \citep{slim2014solutal,depaoli2017solute}.
Experimental measurements in these configurations \citep{neufeld2010convective,backhaus2011convective,depaoli2020jfm,brouzet2022co}, however, suggest a different qualitative and also quantitative behaviour compared to the corresponding Darcy simulations, with a non-linear scaling of $\Sh$ with $\ra^*$.
This discrepancy is likely due to non-Darcy effects \citep{liang2018effect}, i.e. to the pore-induced flow dynamics not captured by Darcy simulations.
Therefore, resolving the flow and the solute transport at the pore-scale is crucial to make reliable models to incorporate in large-scale simulations and to predict the underground fluids migration and mixing, and it represents the main motivation for this work.

When a fluid flows through a matrix of solid obstacles, the fluid follows a random-walk-type path \citep{woods2015flow}.
If the fluid is carrying a solute, in addition to molecular diffusion, solute spreading may occur due to pore-scale change of flow direction (mechanical dispersion), heterogeneities in the aquifer (large-scale dispersion) or other mechanisms, such as dead-end pores (anomalous dispersion).
In this work, we will only refer to mechanical dispersion, i.e. we assume that the medium is homogeneous and without dead-end spaces.
On the one hand, mechanical dispersion produces additional spreading of solute and can be several orders of magnitude more effective than molecular diffusion \citep{delgado2007longitudinal}.
It has been also observed \citep{eckel2022spatial} that the presence of the finger pattern and the counter-current flow structure enhance the longitudinal spreading of the solute compared to unidirectional dispersion of a single-solute plume.
This additional spreading is responsible for the reduction of the local density gradients, diminishing the strength of convective motions and coupling convection and diffusion as mechanisms controlling the mixing process.
To quantify the relative importance of pore structure, material properties and driving force on the overall heat or mass transport process, pore-resolved convective flows have been recently employed in the framework of thermal Rayleigh-B\'enard convection, in both experiments and simulations. 
\citet{chakkingal2019numerical} observe that the flow structure and the heat transfer coefficient are determined by the relative size of thermal length scale (boundary layer thickness) and porous length scale (average pore space).
These properties determine the penetration of the plumes in the boundary layer region, which is responsible for the heat or mass transfer rate. 
In a complementary work, \citet{ataei2019experimental} observed that while at low Rayleigh numbers the transport mechanism is less efficient than in free fluids Rayleigh-B\'enard convection, at larger Rayleigh numbers the classical $\nus$ vs. $\ra$ scaling \citep{grossmann2000scaling,grossmann2001thermal} is recovered. 
The nature of this transition has been investigated by \citet{Liu2020}.
They used two-dimensional direct numerical simulations to examine in detail the microscale flow field through a bead pack.
They observed that the transition between these two regimes is controlled by two physical mechanisms induced by the porous matrix:
(i) the presence of obstacles makes the flow more coherent, with the correlation between temperature fluctuation and vertical velocity enhanced and the counter-gradient convective heat transfer suppressed, leading to heat transfer enhancement; and 
(ii) the convection strength is reduced due the impedance of the obstacle array, corresponding to heat transfer reduction. 
They observed that the scaling crossover occurs when the thickness of the thermal boundary layer is comparable to the averaged pore length scale. 
In addition to the porous structure and the Rayleigh number, in case of thermal convection, the boundary layer thickness and the heat transfer coefficient are determined also by the value of thermal conductivity of the solid and liquid phases \citep{korba2022effects,zhong2023thermal}.

The discussed findings refer to systems in which the medium is permeable to the scalar (heat or solute) transported.
In the frame of solute dispersion, two major differences arise compared to thermal convection, namely the solid is usually impenetrable (over the flow time scales) to the solute, and the ratio between momentum and solute diffusivity (Schmidt number, $\sch$) is typically about two orders of magnitude larger than the ratio between momentum and heat diffusivity (Prandtl number, $\pr$). 
\citet{gasow2020,gasow2021macroscopic} focused on solute convection  investigating solute transport at the pore-scale in Rayleigh-B\'enard configuration, with the aim of deriving corrections to be included in Darcy models to account for the obstacles-induced solute dispersion. 
They observed that the pore-induced dispersion, which may be as strong as buoyancy, affects also the momentum transport and it is determined by two length scales (the pore length scale and the domain height).
Moreover, the dissolution coefficient ($\Sh$) is observed to depend also on the Schmidt number \citep{gasow2022prediction}, in addition to the Rayleigh number and the pore structure \citep{Liu2020}.
These numerical studies, which represent a fundamental step to make large-scale predictions of dissolution in porous media, are still computationally limited to two-dimensions, moderate Rayleigh numbers and large porosity values.
In comparison, experiments can achieve larger Rayleigh numbers and values of porosity that are more representative of geological formations.
In contrast, in the instance of solutal convection, the Rayleigh-B\'enard and the semi-infinite configurations are hard to tackle experimentally because of the limitations associated with keeping the solute concentration constant at the boundaries. 
The porous Rayleigh-Taylor instability, relatively less studied compared to the corresponding Rayleigh-B\'enard and one-sided configurations, is an excellent candidate to overcome this obstacle, and it is the object of this work.

A porous Rayleigh-Taylor system consists of two miscible fluids of different density initially arranged in an unstable configuration and immersed in a porous matrix.
After an initial diffusive phase \citep{Wang2016,wang2018effect}, the flow is driven by convection, and due to the transient nature of this system, the mixing rate of the two fluids varies in time. 
In geophysical applications, it is crucial to determine the mixing rate of the involved fluids, to be able to make reliable predictions on the evolution of the volume of subsurface flow.
The porous Rayleigh-Taylor system has been studied at the Darcy scale with the aid of Hele-Shaw cells and Darcy simulations \citep{de2020chemo}.
The flow behaviour is quantified by the mixing length, i.e. the vertical extension of the mixing region.
The growth rate of the mixing length is controlled by the combined action of buoyancy (density difference) and drag (viscous dissipation through the medium) \citep{boffetta2022dimensional,depaoli2022experimental}.
In the case of miscible fluids, the role of diffusion across the fluid-fluid interface is also crucial \citep{gopalakrishnan2017relative}, as it weakens convection by reducing local density gradients.
As a result of this time-dependent finger growth process, the mixing dynamics is also transient, and it may be quantified via the mean scalar dissipation rate \citep{hidalgo2012scaling}.
The mean scalar dissipation is particularly convenient because it can be exactly related to other flow quantities, e.g. the averaged concentration fluctuations, and it has been already described numerically at the Darcy scale \citep{depaoli2019universal,depaoli2023convective}.
Also in the Rayleigh-Taylor case, however, the behaviour of dispersion cannot be captured by Darcy simulations despite its crucial role, as it can influence dramatically the onset of the gravitational instabilities \citep{menand2005dispersion} and the mixing dynamics.
In this flow configuration, the full flow dynamics has been studied with the aid of three-dimensional simulations by \citet{sardina2018buoyancy}, where the authors consider a thermal convection at relatively large values of porosity ($0.6-1$).
They proposed a model to incorporate the effect of the medium within a friction coefficient to be included in the Navier-Stokes equations.
The case of solutal convection in geological formations may be markedly different, since the porosity is low (0.2 -- 0.4) and the Schmidt numbers about two orders of magnitude larger than in the thermal case, and we aim precisely at this gap.

We investigate a Rayleigh-Taylor instability in a saturated, homogeneous and isotropic porous medium.
We present the results from Hele-Shaw type experiments with bead packs and two-dimensional numerical simulations, where we resolve the flow at the pore-scale at high Rayleigh-Darcy numbers, high Schmidt numbers, and low values of porosity.
Experiments and simulations are specifically designed to reproduce the same fluid and medium properties, namely linear density-concentration dependency, matching porosities, and porous medium impermeable to the transported scalar (solutal convection).
In the experiments, porous media and fluid properties are varied, and different flow regimes are observed, namely a Darcy-type flow (with the flow structures larger than the pore length scale) and a diffusion-dispersion regime, with the strength of mechanical dispersion being equivalent or dominant with respect to molecular diffusion.
Simulations are first validated against the experimental measurements in terms of evolution of the mixing length, and then used to quantify the evolution of the mixing rate, measured by the mean scalar dissipation. 
Several dissolution regimes have been identified: Initially, the flow is controlled by diffusion.
This regime is followed by a convection-dominated phase.
The average concentration profiles follow a self-similar behaviour, which we describe theoretically throughout the flow dynamics.
In the final regime finite-size effects reduce the solute transport.
Our experimental and numerical results are used to explain the evolution of the flow with the aid of physically grounded models.

Our findings are relevant to the convection of miscible fluids in porous media. 
However, we note that some differences occur when CO$_2$ sequestration is considered, including the non-monotonic relationship between the fluid density and the solute concentration \citep{hidalgo2015dissolution}, and the partial miscibility of the phases involved \citep{huppert2014fluid}. 
These effects, as well as additional limitations (dimensionality of the systems and idealized structure of the media) later discussed in detail, prevent a direct application of the present findings to CO$_2$ dissolution in geological formations.
Nonetheless, these results represent a fundamental ingredient required to build a modelling framework for large-scale simulations, as they are obtained in a well-defined and controlled physical system.

The paper is organised as follows.
In \S\ref{sec:meth2} the problem is formulated.
We describe the experimental and the numerical setups in \S\ref{sec:meth1}. 
We present our results in terms of qualitative flow dynamics (\S\ref{sec:results}), mixing length and concentration profiles (\S\ref{sec:results2}), flow structure and wavenumber (\S\ref{sec:results3}).
In \S\ref{sec:results4} we quantify the dissolution rate, which we describe by physical models in each of the regimes identified.
Finally, in \S\ref{sec:concl} we summarise our findings and provide a brief perspective on future research directions.

\section{Problem formulation}\label{sec:meth2}

\begin{figure}
\centering
\includegraphics[width=0.95\columnwidth]{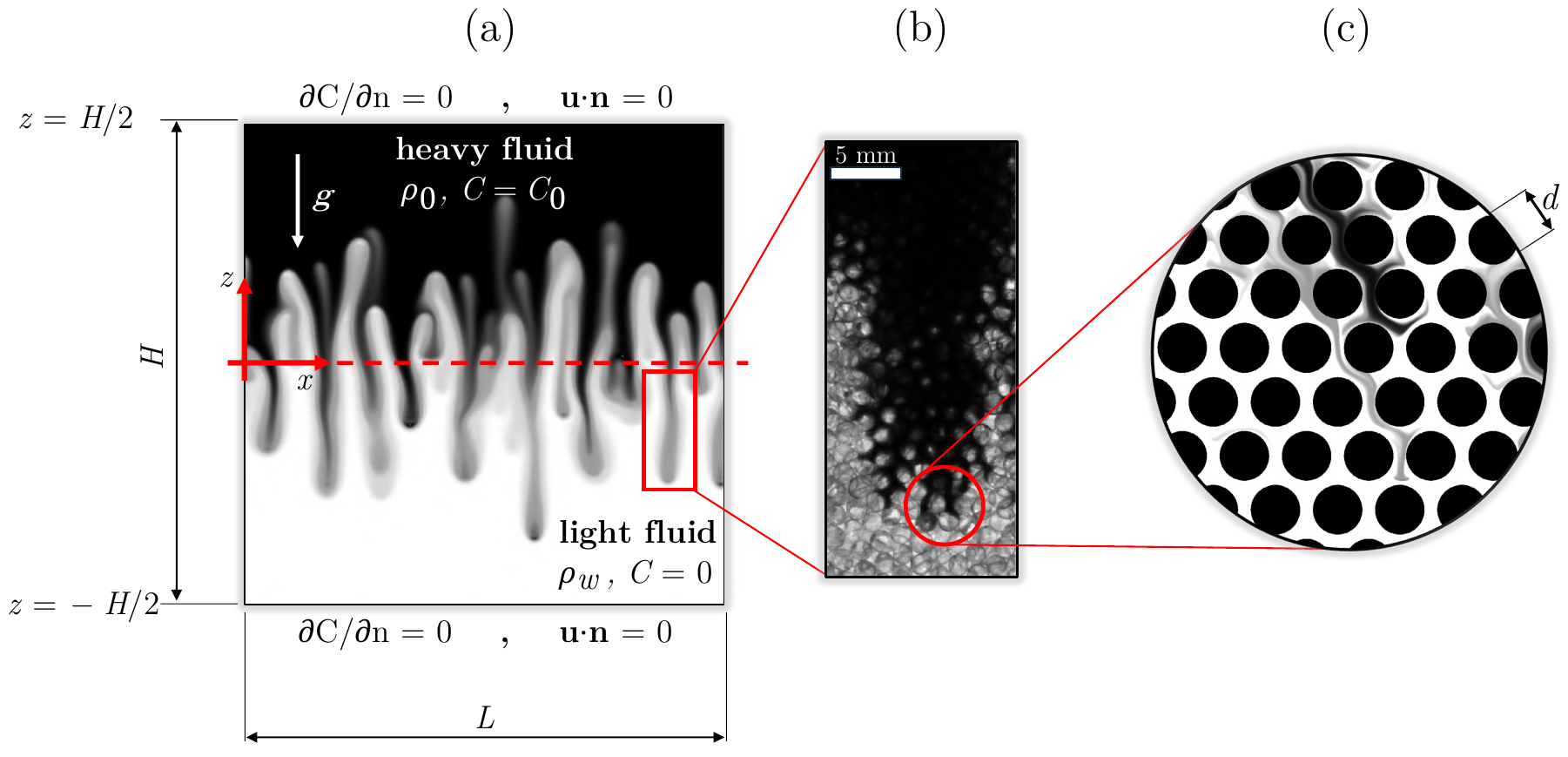}
\caption{
Sketch illustrating the system considered.
(a)~Domain with explicit indication of the boundary conditions (no flux of mass or solute through the horizontal walls) and domain dimensions in horizontal ($L$) and vertical ($H$) directions.
The reference frame ($x,z$) as well as the initial position of the interface (red dashed line) are indicated, with the heavy fluid (density $\rho_0$, concentration $C=C_0$) initially lying on top of the lighter one ($\rho_w$, $C=0$).
(b)~In the experiments, a transparent medium consisting of monodisperse beads and fluids of different colours are used.
(c)~In the simulations, the geometry consists of an array of spheres (diameter $d$) fully saturated with fluid. 
The fluid carrying the solute moves thought the spheres.
}\label{fig:sk} 
\end{figure}

The process of convective dissolution is studied here in the frame of the Rayleigh-Taylor instability.
It can be modelled as two layers of miscible fluids having different density, initially separated by a flat interface, and located in an accelerated reference frame \citep{boffetta2017incompressible}. 
The process is simulated in the context of porous media flows, mimicked with the aid of a porous layer (height $H$, width $L$) made of spheres (diameter $d$). 
The fluids fully saturate the medium, have same viscosity ($\mu$) but different density, and are arranged in an unstable configuration, with the heavy fluid (density $\rho_0$) lying on top of the lighter one (density $\rho_w$).
Therefore, the maximum density difference within the system is $\Delta\rho =\rho_0-\rho_w$.
The system, consisting of a porous slab saturated with two fluids in an accelerated field (gravity acceleration, $g$), is sketched in figure~\ref{fig:sk}(a). 
The density difference is induced by the presence of a solute, which is quantified by the solute concentration $C$, taking its maximum at the upper layer ($C=C_0$) and minimum at the lower layer ($C=0$).
The reference frame ($x,z$) is defined as in figure~\ref{fig:sk}(a) such that it is centred at the mid height of the domain and $z$ is aligned with $g$ but in opposite direction.
We aim at mimicking a system with horizontal boundaries ($z=\pm H/2$) that are impermeable to both fluid and solute, and we consider
\begin{equation}
\mathbf{u}\cdot\mathbf{n}=0\quad,\quad \frac{\partial C}{\partial n}=0 \text{  ,}
\label{eq:rt_ra0exp4}
\end{equation}
with $\mathbf{n}$ the unit vector perpendicular to the boundary.

In the present flow configuration, the dimensionless parameters controlling the system can be grouped in three-main categories: medium parameters (Darcy number, porosity), fluid parameters (Schmidt number) and flow parameters (Rayleigh, Rayleigh-Darcy, Peclet and Reynolds numbers). 

We consider a simplified configuration in which the medium is homogeneous and isotropic.
Assuming the structure obtained from the sphere packing as an isotropic and homogeneous medium, it can be fully described by two global quantities, namely porosity and permeability.
The porosity, $\phi$, represents the ratio between the volume of fluid and the total volume (fluid + solid) of the domain considered, and therefore it varies between $\phi=0$ (pure solid) and $\phi=1$ (pure fluid). 
The permeability, $k$, quantifies the ability of the porous matrix to allow a fluid to flow through it.
For a given geometry of the medium, the Darcy number 
\begin{equation}
\da=\frac{k}{H^2}
\label{eq:meth01} 
\end{equation} 
quantifies the relative dimension of the microscopic pore-scale ($\sqrt{k}$) and the macroscopic length-scale ($H$) \citep{hewitt2020vigorous}.

The dimensionless parameter that quantifies the fluid properties is the Schmidt number, which is the ratio of momentum diffusivity (kinematic viscosity, $\mu/\rho_o$) to mass diffusivity $D$,
\begin{equation}
\sch=\frac{\mu}{\rho_0 D}\text{ .}    
\label{eq:sc}
\end{equation}

The dimensionless flow parameters and the relevant flow scales are obtained by combining domain, medium and fluid properties.  
A possible velocity scale of the flow is the buoyancy velocity 
\begin{equation}
U=\frac{g\Delta\rho k}{\mu}\text{  ,}
\label{eq:buoy_velocity} 
\end{equation} 
which is obtained at the equilibrium between driving forces ($gk\Delta\rho$) and viscous dissipation through the medium ($\mu$).

Multiple length scales are effective in this problem.
One can consider as a reference length scale the distance $\ell$ over which advection and diffusion balance \citep{slim2014solutal}
\begin{equation}
\quad\ell=\frac{\phi D}{U}\text{  .}
\label{eq:eq34} 
\end{equation} 
Possible alternatives consists of the characteristic bead size (sphere diameter, $d$) or the domain height ($H$).
We will see that each of these scales is relevant in different phases of the dissolution process.
Solutal convection in pure fluids is characterized by the competing effect of convection (solute-induced density differences) and dissipation or diffusion, respectively.
The relative importance of these contributions is measured by the concentration Rayleigh number based on the domain size ($\ra$) or on the diameter of the spheres ($\ra_d$),
\begin{equation}
\ra=\frac{g\Delta\rho H^3}{\mu D}\quad\text{,}\quad
\ra_d=\frac{g\Delta\rho d^3}{\mu D}=\ra\left(\frac{d}{H}\right)^3\text{ ,}
\label{eq:eq35} 
\end{equation} 
respectively. 
These parameters include convection and dissipation, but do not consider the presence of the medium, which has a stabilizing effect on convection due to the additional friction on the surface of the pores.
The ratio of the strength of these contributions is estimated by the Rayleigh-Darcy number
\begin{equation}
\ra^*=\frac{g\Delta\rho k H}{\mu\phi D}=\frac{UH}{\phi D}=\frac{H}{\ell}=\frac{\ra\da}{\phi}.
\label{eq:eq36} 
\end{equation} 
We remark that the concentration Rayleigh number \eqref{eq:eq35} and the Rayleigh-Darcy number \eqref{eq:eq36} are linked to the porous medium properties via the Darcy number \eqref{eq:meth01} and the porosity.
Finally, two more flow parameters are used to determine whether the flow can be modelled as a Darcy flow or not. 
Following \citet{hewitt2020vigorous}, the flow can be considered as a Darcy-type flow, if the length scale of the flow structures is much larger than the representative volume over which the quantities are averaged.
It is obtained for (i)~viscous forces dominating over inertia at the pore-scale, and (ii)~length scale of the convective flow large compared to the pore size.
These conditions are fulfilled if
\begin{equation}
\Rey=\frac{\ra^*\da^{1/2}}{Sc}\ll 1 \quad,\quad
\pe=\ra^*\da^{1/2}\ll 1,   
\label{eq:eqpere} 
\end{equation}
with $\Rey$ and $\pe$ the pore-scale Reynolds number and the Peclet number, respectively.
In this study, only a few experiments (and no simulations) fall in the Darcy case, and the relative flow dynamics will be discussed later in \S\ref{sec:results}.
Note that in this definition of $\pe$ it is assumed that the pore-scale length used as a length-scale for $\pe$ is $\sqrt{k}$.
An alternative choice consists of using $d$, which would produce larger values of $\pe$ (by a factor of approx.~7.5 in this configuration).

The flow scales of the experiments are listed in table~\ref{tab:listex2}, while the dimensionless parameters corresponding to present experiments and simulations is reported in table~\ref{tab:listex3}.

\section{Methodology}\label{sec:meth1}
We investigate convective mixing in confined porous media. 
The flow driving force consists of the density differences induced by the presence of a solute.
We consider two miscible fluids characterized by a linear dependency of density with concentration.
The fluids are immersed in a fully saturated, homogeneous and isotropic porous medium. 
Laboratory experiments and numerical simulations are used to investigate the problem.
In \S\ref{sec:meth3a} we present the experimental setup, consisting of a bead pack and an optical measurement system.
Pore-resolved two-dimensional simulations, where circular impermeable obstacles are employed to mimic the solid matrix of the porous medium, are discusses in \S\ref{sec:meth4}.

\subsection{Laboratory Experiments}\label{sec:meth3a}
The laboratory experiments are performed with the aid of a thick Hele-Shaw cell filled with monodisperse beads and saturated with two fluids of different densities in an unstable configuration.
The parameters that can be varied are the density difference $\Delta\rho$ between the fluids, and the diameter $d$ of the beads. 
Combining these parameters one can determine the flow reference scales, namely $\ell$ and $U$.
The experiments performed are listed in table~\ref{tab:listex2}.

\begin{figure}
    \centering
    \includegraphics[width=0.95\columnwidth]{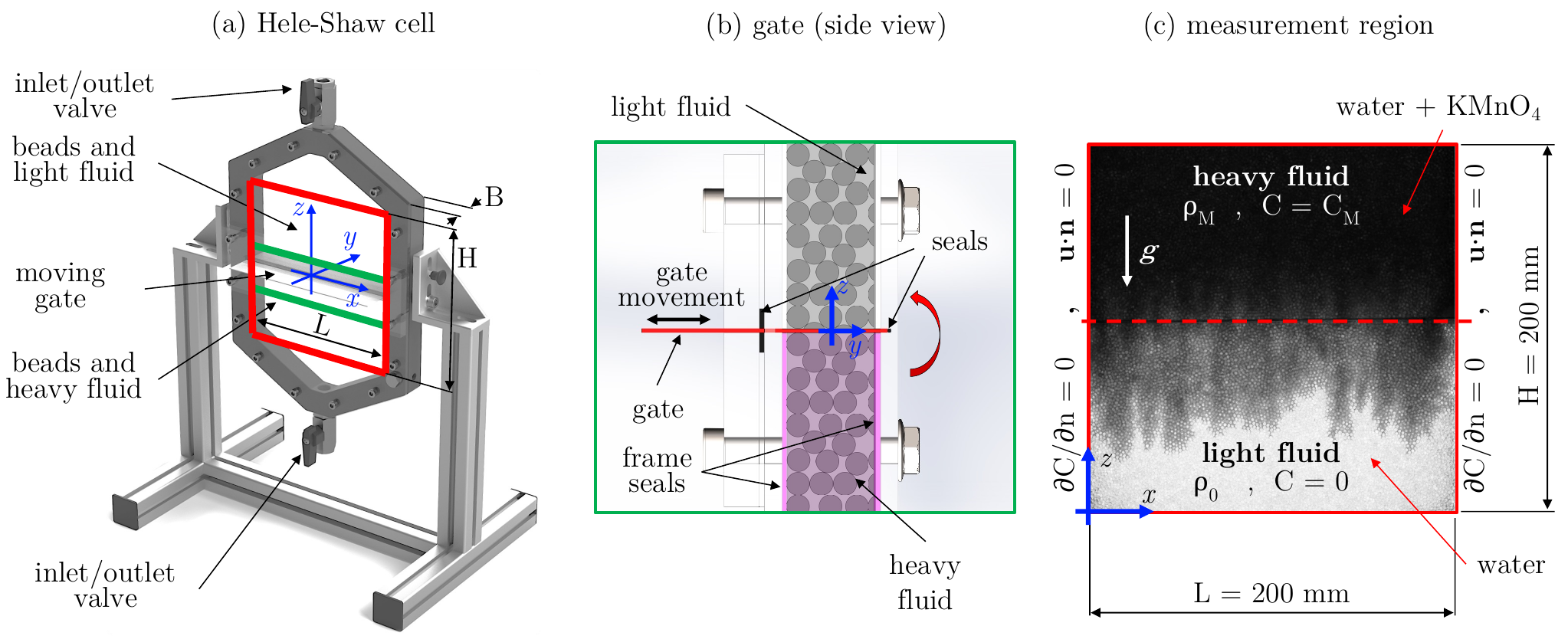}
    \caption{
    (a)~Schematic representation of the Hele-Shaw cell. 
    The cell is filled with beads and different fluids from the upper and lower valves. 
    The gate keeps the fluids initially separated. The reference frame of the lab ($x,y,z$) is shown. 
    The area of the gate (highlighted in green) and the measurement region (highlighted in red) are discussed in detail in the panels (b) and (c).
    (b)~Side view of the gate mechanism. 
    The gate is removed and the fluids can mix. 
    A system of seals is used to prevent mixing of the fluids before the gate is open. 
    After the gate is extracted, the cell is rotated about the $x$ axis (i.e., upside down) as indicated by the red arrow.
    (c)~Front view of the measurement region. After the cell is rotated, the fluids are in an unstable configuration (heavy solution on top of the lighter fluid) and the recording starts. 
    }
    \label{fig:fig2}
\end{figure}

In the following, we will discuss the experimental setup, the measurement procedure, the fluid properties and the porous medium properties.
The employed Hele-Shaw cell used is sketched in figure~\ref{fig:fig2}(a).
It consists of a hexagonal container with uniform thickness. 
The shape is designed to simplify the fluid extraction/injection phases.
The hexagonal walls (made of PMMA layers, thickness 8~mm) are transparent to allow optical access to the flow, and are kept in place by a thick frame (25~mm) and bolts.
A gasket (2~mm thick) is placed between the frame and the walls to prevent fluid leakage.
The cell is initially filled with monodisperse glass spheres, then it is vibrated to consolidate the medium, and finally the two fluids of different density are injected via the upper and lower valves.
Note that a gate (thickness 1 mm, material polyethylene terephthalate - PET), as indicated in the side view in figure~\ref{fig:fig2}b), initially divides the upper and the lower sides of the cells to prevent fluids from mixing.
A 1~mm high housing, equipped with a rubber seal all along its length, is obtained on the front wall, and the gate fits in.
The gate can slide through an opening located on the back side of the cell, which has additional seals hold in place by a plastic frame.

When the medium is consolidated and the fluids are injected, the cell is placed in a stable configuration (with light fluid on top of the heavy one).
The gate is later removed and the cell rotates about the $x$ axis (red arrow in figure~\ref{fig:fig2}b) to turn in the initial unstable condition (heavy fluid on top of the light one) chosen to initialize the experiment.
The measurements are performed only in the central portion of the domain, indicated in red in figure~\ref{fig:fig2}(a) and discussed in figure~\ref{fig:fig2}(c), which has size $H\times L \times B=200\times200\times 28.5$~mm$^3$.
The cell is illuminated on one side by a LED light (Phlox LEDW-BL $300\times300$~mm$^{2}$) and on the opposite side a high-resolution camera (Nikon D850 with lenses AF Nikkor, 50~mm, 1:1.8D) records the evolution of the flow at an acquisition rate that varies between 0.05 and 2 frames per second.

To allow here a reliable comparison of experimental results against the direct numerical simulations, the fluids employed in the experiments have been specifically selected because of the linear dependency of the density of the solution with respect to the solute concentration \citep{slim2013dissolution,depaoli2022experimental}.
This condition is indeed met not only in the simulations presented here, but also in most of the numerical works available in the literature.

The employed fluids are water and an aqueous solution of KMnO$_4$ (Potassium Permanganate, Thermo Scientific, ACS reagent).
We consider that the dynamic viscosity, $\mu=9.2\times10^{-4}$~Pa$\cdot$s, is constant and independent of the solute concentration \citep{slim2013dissolution}.
Similarly, we assume that the diffusion coefficient is not sensibly affected by solute concentration, and corresponds to $D=1.65\times10^{-9}$~m$^{2}$/s.
While water density ($\rho_w$) is nearly constant among the experiments (it is only dependent on temperature, $\vartheta$), the density of the aqueous solution of KMnO$_4$ can be varied by changing the solute concentration in the aqueous solution, $C$, which is used to control the density difference between the heavy and the light fluid. 
The mass fraction of the solution is defined as 
\begin{equation}
\omega(C) = \frac{C}{\rho(C)}\text{  .} 
    \label{eq:defmf}
\end{equation}
The respective dependency of density, mass fraction and concentration can easily be determined. 
The density of the mixture, $\rho(C)$, can be well approximated by a linear function of the solute concentration, i.e., it meets the desired condition:
\begin{equation}
\rho=\rho_0\biggl[1+\frac{\Delta\rho }{\rho_0C_0}\bigl(C-C_0\bigr)\biggr]\textit{  .}
\label{eq:rt_eq4a}
\end{equation}
The concentration-density profiles as well as additional details are reported in Appendix~\ref{sec:appa1}.

With the aim of mimicking a homogeneous and isotropic porous medium, we fill the cell with monodisperse spheres having diameter $d$, with 1~mm $\le d \le 4$~mm.
Provided that the spheres are monodisperse, the diameter of the beads and the porosity of the medium are the two parameters that determine the medium property, i.e., the permeability.
In the following, we will discuss bead size, medium porosity and permeability.
Again, a summary of all the experimental parameters considered is reported in table~\ref{tab:listex2}.

The porosity of the medium indicates the ratio between the volume of fluid used to fill the cell and the total cell volume (fluid and beads).
We measure the porosity by comparing the volume of water required to fill the cell with and without the beads. 
The preparation of the medium is crucial in determining the cell porosity and permeability.
In this work, the cell is vibrated before injecting the fluid so that the medium consolidates.
Following this procedure, the beads form a close random packing and the expected value of porosity in case of monodisperse spheres is $\phi = 0.359 - 0.375$ \citep{dullien2012porous,haughey1969structural}.
The values of porosity measured experimentally are $\phi=0.37$ for all nominal diameters considered, except $d=3.00$~mm, in which the value of porosity measured is slightly lower ($\phi=0.35$).
This difference can be possibly attributed to the lower quality of the beads with $d=3.00$~mm, which have a wide distribution of diameters (see Appendix~\ref{sec:appa0}).
Indeed, the more dispersed the diameters, the lower the value of porosity that can be achieved.

\begin{table}
\begin{center}
\def~{\hphantom{0}}
\setlength{\tabcolsep}{5pt}
\footnotesize
\begin{tabular}{@{}ccccccccccccccc@{}}
Name && \multicolumn{3}{c}{Fluid} && \multicolumn{3}{c}{Medium} && \multicolumn{2}{c}{Flow} \\
\midrule
& & $\vartheta$ & $\omega_M$ &  $\Delta\rho$ & & $d$ & $\phi$ & $k$  & & $\ell$ & $U$  \\
& & {[}$^\circ$C{]} & {[}-{]} & {[}kg/m$^3${]} & & {[}mm{]} & {[}-{]} & {[}m$^2${]} &  & {[}m{]} & {[}m/s{]} \\\\
\input{table_experiments_dim}
\end{tabular}
\caption{\label{tab:listex2}
Dimensional parameters of the performed experiments. 
The domain size is constant ($L=H=200$~mm, $B=28.5$~mm). 
Fluid temperature ($\vartheta$), solute mass fraction ($\omega_M$), and density contrast ($\Delta\rho$) are reported.
Fluid viscosity and diffusivity are $\mu=9.2\times10^{-4}$ Pa$\cdot$s and $D=1.65\times10^{-9}$~m$^2$/s, respectively. 
Water density is assumed $\rho_0=10^3$~kg/m$^3$.
Bead diameter ($d$), medium porosity ($\phi$), and permeability ($k$) are indicated.
Flow length scale $\ell$ and velocity scale $U$, defined as in \eqref{eq:eq34} and \eqref{eq:buoy_velocity} respectively, are also reported.
}
\end{center}
\end{table}

The permeability is inferred from the Kozeny-Carman correlation, i.e. 
\begin{equation}
k=\frac{d^2}{36k_C}\frac{\phi^3}{(1-\phi)^2}
\label{eq:eqck}
\end{equation}
where $k_C$ is the Carman constant. 
This phenomenological correlation is obtained for creeping flow, and it is found to be independent of the particle shape \citep{dullien2012porous} (for non-spherical particles, $d$ is the equivalent diameter).
The Carman constant originally proposed within the Kozeny-Carman formulaton (monodisperse spheres) is $k_C = 4.8\pm0.3$, usually approximated to 5, which gives $36k_C=180$. 
At a later time, Ergun proposed the Blake-Kozeny formulation, in which the coefficient is $36k_C=150$. 
Both formulations are based on fitting of experimental data for flow through granular beds.
Other formulations have been proposed to improve the predictions at low or high values of porosity, as well as to include the effect of the Reynolds number.  
A detailed review on the possible values of $k_C$ is provided by \citet{xu2008developing}, where a more general phenomenological formulation is also proposed. 
For the specific case of monodisperse spheres randomly packed, \citet{zaman2010hydraulic} have shown that the Kozeny-Carman formulation~\eqref{eq:eqck} with $k_C=5$ provides good results, and this is the correlation we employ.

\begin{figure}
    \centering
    \includegraphics[width=0.92\columnwidth]{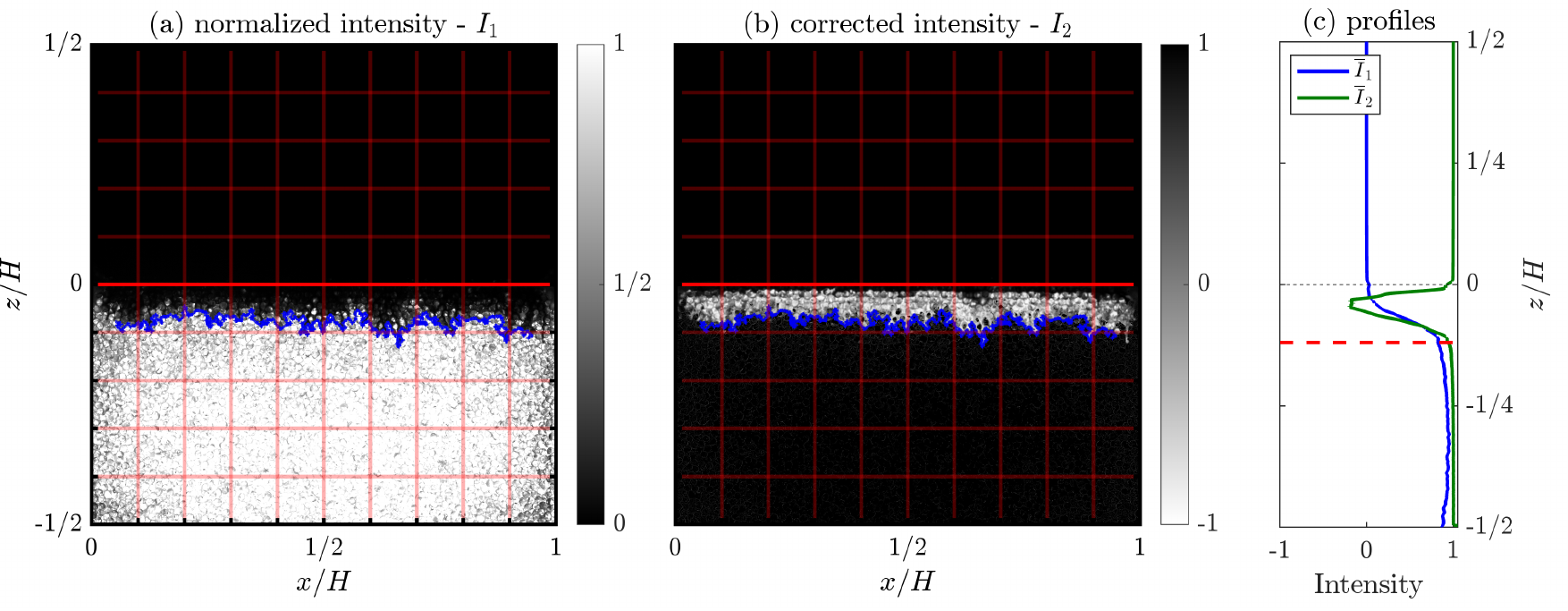}
    \caption{Processing of images. 
    (a)~Normalised intensity profile $I_1$, defined in equation~\eqref{eq:norm1}, shown over the entire measurement region for the experiment E12.
    The initial position of the interface ($z/H=0$, red solid line) is reported.
    The instantaneous position of the interface, defined as an iso-contour of $I_{1}$, is also shown (blue line).
    (b)~Corrected intensity profile $I_2$, defined as in equation~\eqref{eq:norm2}.
    The corrected intensity $I_{2}$ is lower than 1 within the mixing region.
    (c)~Horizontally-averaged intensity profiles. 
    The lower edge of the mixing region, determined by the position where $\overline{I}_{2}$ is lower than a given threshold value, is reported (dashed red line).
    This is related to the mixing length $h$, discussed in detail in \S\ref{sec:results2}.
    See also the supplementary material (movie 1).
    }
    \label{fig:int3}
\end{figure}

The flow is recorded by the camera.
The raw images of the measurement region are analysed to quantify the flow evolution.
An example is shown in figure~\ref{fig:int3}, where an instantaneous intensity field obtained from experiment E12 is discussed. 
The measurement region (see figure~\ref{fig:fig2}c) consists of the central squared portion of the cell, having size $H\times H = (200$~mm$)^{2}$.
The denser solution, initially on top, is much darker than the lighter one (transparent). 
The thickness of the cell is large, and the light intensity detected at the upper layer is weakly evolving during the experiment.
Therefore, for better accuracy and due to symmetry of the system, we consider only the lower portion of the cell for the experimental measurements ($z/H\le0$).

The pre-processed fields are analyzed to produce intensity profiles and infer the time-dependent evolution of the interface.
The pre-processing consists of several steps, illustrated in figure~\ref{fig:int3}. 
The initial light intensity distribution $I(x,z,t=0)$, obtained from the raw images, is used to compute and store the initial mean light intensity values of the high ($I_H$, $z/H>0$) and low ($I_L$, $z/H<0$) fluid density layers.
The normalised light intensity field 
    \begin{equation}
        I_1(x,z,t) = \frac{I(x,z,t)-I_H}{I_L-I_H}.
        \label{eq:norm1}
    \end{equation}
is suitable to visualise the instantaneous flow configuration. 
 It is used, for instance, to identify the instantaneous position of the interface of the mixing region (figure~\ref{fig:int3}a).
However, we observe in figure~\ref{fig:int3}(c) that the horizontally-averaged intensity profile $\overline{I}_1$ varies smoothly within the mixing region, and therefore it is not a good indicator to determine the edge of the interface.
We introduce the corrected intensity (figure~\ref{fig:int3}b), $I_2$, defined as
     \begin{equation}
        I_2(x,z,t) = 2\left [\frac{1}{2} + I_1(x,z,t) - \widehat{I}_1(x,z,0) \right] ,
        \label{eq:norm2}
    \end{equation} 
with $\widehat{I}_1(x,z,0)$ the initial intensity field $I_1(x,z,t=0)$ computed with a moving average (squared window of size 10~pixel).
We observe that $I_{2}$ is lower than 1 only within the mixing region and this property, which is also clear from the horizontally-averaged profiles in figure~\ref{fig:int3}(c),  makes $I_2$ a more reliable observable to quantify the extension of the mixing region (red dashed line).

\subsection{Numerical simulations}\label{sec:meth4}

\begin{figure}
    \centering
    \includegraphics[width=0.95\columnwidth]{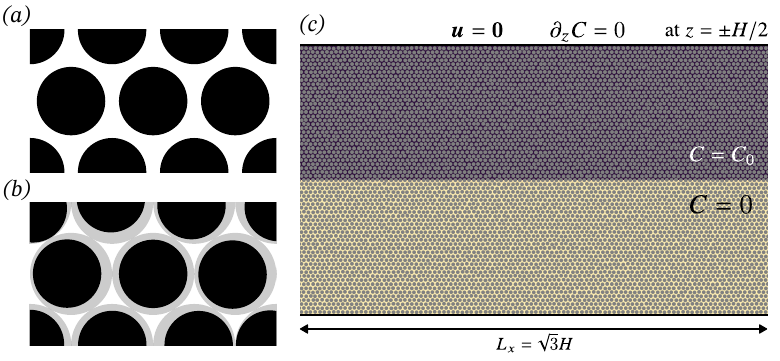}
    \caption{
    Schematic of the porous medium layout in the simulations.
    $(a)$ Regular hexagonal distribution of solid circles used in the simulations.
    $(b)$ An example of random perturbations to the regular pattern. Grey areas mark the possible area in which each solid circle can move without overlapping another.
    $(c)$ Schematic of the total domain for $H/d=70$, with solid particles in grey, and the fluid domain coloured according to the initial concentration.
    }
    \label{fig:pore_sim_schematic}
\end{figure}

In the numerical part of this study, we solve the Navier--Stokes equations for momentum, subject to the Oberbeck--Boussinesq approximation.
This means that variations in the density are only significant in the buoyancy term.
We assume the flow to be incompressible and impose $\grad \cdot \bu = 0$ on the velocity field $\bu$.
We consider variations in density to be linearly related to the concentration field $C$, which itself satisfies an advection-diffusion equation.
We therefore consider the following governing equations
\begin{align}
    \partial_t \bu + (\bu \cdot \grad)\bu &= -\rho_0^{-1}\grad p + \nu \nabla^2 \bu - g\beta C \hat{\boldsymbol{z}}, \label{eq:NSE1} \\
    \partial_t C + (\bu \cdot \grad)C &= D \nabla^2 C, \label{eq:NSE2}
\end{align}
where $\rho_0$ is a reference density, $\nu=\mu/\rho_0$ is the kinematic viscosity, $D$ the solutal diffusivity, $g$ gravitational acceleration, and $\beta$ the solutal contraction coefficient describing the linear relationship between density and concentration.
The pressure $p$ satisfies a Poisson equation such that a divergence-free velocity field is ensured.

We solve the equations \eqref{eq:NSE1}-\eqref{eq:NSE2} in a two-dimensional domain of height $H$ and of width $\sqrt{3}H$.
Periodic boundary conditions are imposed in the horizontal ($x$) direction for all flow variables.
At the top and bottom walls ($z=\pm H/2$), we impose the boundary conditions~\eqref{eq:rt_ra0exp4}, i.e. zero mass flux of solute ($\partial_z C = 0$) and no penetration ($w=0$, being $w$ the vertical velocity). 
In addition, since the pore-scale flow is resolved, also the no-slip condition applies along these walls ($\bu=\boldsymbol{0}$).
In the following subsections, we describe the properties of the solid porous matrix that occupies a portion of the simulation domain, as well as details of the numerical implementation for the flow solver and the conditions at the liquid-solid boundaries.

The numerical simulations are designed to match the porosity of the experiments, namely $\phi=0.37$.
To consider a two-dimensional setup most similar to the monodisperse spherical bead pack of the experiments, we construct the solid phase in the simulations from circles of a given diameter $d$.
Most past studies of a similar configuration \citep[e.g.][]{sardina2018buoyancy,Liu2020} that explicitly resolve the pore-scale dynamics with a liquid-solid boundary are performed at a higher porosity, allowing for a wide range of configurations for the solid phase.
Since we aim to match a low experimental porosity of $\phi=0.37$, we prescribe a hexagonal arrangement of the circular beads, as shown in figure \ref{fig:pore_sim_schematic}(a), which allows for free percolation of the fluid in 2D at these low porosities.
Such perfectly regular arrangements are not representative of the porous matrix in the experiments, so we also repeat our simulations in domains in which random shifts from this hexagonal arrangement are made to the positions of the solid circles.
An example of these random shifts is shown in figure \ref{fig:pore_sim_schematic}(b).
To prevent regions of trapped fluid, we limit the random perturbations to the grey areas in that schematic, such that the (black) solid regions do not overlap.

As discussed above in \S\ref{sec:meth3a}, the permeability $k$ is key to understanding the effect of the porous medium properties on the flow.
For example, the velocity scale $U$ of \eqref{eq:buoy_velocity} relies on the balance between buoyancy, permeability, and viscosity.
While the determination of the permeability for three-dimensional arrays of spheres is well studied, for two-dimensional flows (array of cylinders with infinite length) the situation has been investigated less.
By definition, the value of permeability would be determined by measuring the pressure drop across the medium for different flow rates.
\citet{happel2012low} suggest that $k_C=5$ holds also for two-dimensional media.
They considered a flow perpendicular to an array of cylinders \citep[indicated as perpendicular flow in][]{xu2008developing} and observed that for $0.25<\phi<0.55$, the Carman constant can very well approximated as $k_C=5$. 
Therefore, also in the two-dimensional case, we will assume that equation~\eqref{eq:eqck} with $k_C=5$ applies.

We use the highly-parallelised AFiD (Advanced Finite-Difference) code to perform our simulations.
The governing equations \eqref{eq:NSE1}-\eqref{eq:NSE2} are solved using second-order central finite differences to compute spatial derivatives, with time-stepping performed using an implicit Crank-Nicolson method for the vertical diffusive terms ($\partial_{zz}$) and a third-order explicit Runge-Kutta scheme for all other terms.
A fractional-step method is used to impose the divergence-free condition at each time step, where a Poisson equation is solved for the pressure.
Further details of the numerical scheme can be found in \citet{verzicco_finite-difference_1996} and \citet{van_der_poel_pencil_2015}.
A multiple-resolution method is applied to the concentration field for accurate simulation at low diffusivities, following \citet{ostilla-monico_multiple-resolution_2015}.
Cubic Hermite interpolation is used to interpolate the concentration field to the velocity grid for the buoyancy forcing, and to interpolate the velocity field to the refined grid for advection of the concentration field.

The solid phase is handled using the immersed boundary method \citep{verzicco_immersed_2023}.
We follow the direct forcing approach of \citet{fadlun_combined_2000} such that zero velocity is imposed in the solid during the implicit step of the numerical solution.
In this approach, linear interpolation is used at the first grid nodes in the fluid from the solid boundary, so that the interface is captured more accurately than the grid resolution.
Similarly for the concentration field, we ensure zero normal gradient at the immersed boundary by specially treating these boundary nodes.

The collection of performed simulations are listed in table~\ref{tab:listex3}.
For each set of parameters, we perform three simulations: one with the regular hexagonal packing defining the centre positions of the solid circles, and two with (distinct) random perturbations to this arrangement.
Each simulation is initialised with zero velocity, and an initial concentration field of
\begin{equation}
    \left.C(z)\right|_{t=0} = C_0\left[\mathcal{H}(z) + A(z) W\right] , \quad
    A(z) = 0.1 \mathrm{sech}^2\left(100\, z/H\right)
\end{equation}
where $\mathcal{H}$ is the Heaviside function, and $W$ is a white noise random variable taking values between $-1$ and $1$, producing a region of random perturbations of width $H/100$ across the interface.
Uniform grid spacing is used in all directions, with a resolution of at least 32 grid points per solid diameter for the velocity grid and a resolution 4 times that for the refined concentration grid.
The largest computational grids are used for simulations S10-S12, where $H/d=70$, with the base grid at a resolution of $4096\times2365$ and the refined grid at a resolution of $16384\times9459$.
This resolution is noticeably higher than in previous studies \citep[e.g.][]{sardina2018buoyancy}, and allows us to accurately simulate the small-scale structures arising from buoyancy-driven flows at high $Sc$.

\begin{table}
\begin{center}
\def~{\hphantom{0}}
\setlength{\tabcolsep}{5pt}
\begin{tabular}{@{} c c c c c c c c c @{}}
Name &  $H/d$ & $\phi$ & $\sch$ & $\ra$ & $\ra_d$ & $\ra^*$ & $\pe$ & $\Rey$ \\\\

\input{table_experiments_JFM}
\\
\input{table_simulations_JFM}
\end{tabular}
\caption{\label{tab:listex3}Dimensionless parameters of the performed experiments (E\#) and the simulations (S\#).
The domain aspect ratio of the experiments and the simulations is 1 and $\sqrt{3}$, respectively. 
}
\end{center}
\end{table}

\section{Flow dynamics}\label{sec:results}
We now present the results of the experiments and simulations.
The experiments are performed by changing the bead diameter ($d$) and the density difference ($\Delta\rho$).
Simulations are performed for different values of diameter-based Rayleigh number $(\ra_d)$ and domain to bead size $(H/d)$.

\begin{figure}
    \centering
    \includegraphics[width=0.9\columnwidth, angle=0]{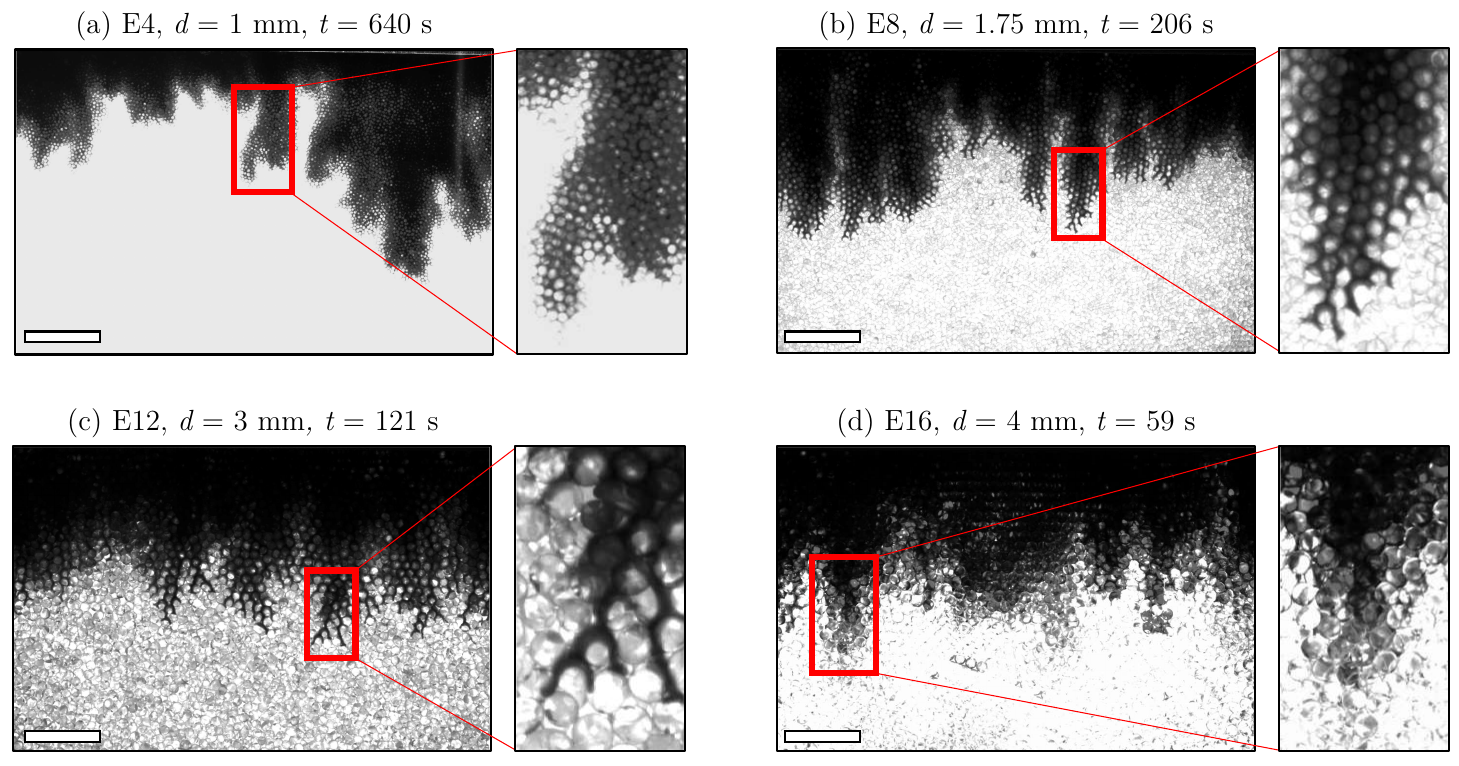}
    \caption{
    Flow evolution (raw images of laboratory experiments) for different bead size, $d$, and same density difference, $\Delta\rho\approx7$~kg/m$^3$.
    The permeability is increased while the driving force remains unchanged.
    The flow structure changes remarkably from (a) to (d), due to the change in the size of the fingers relative to the beads.
    For visualisation purposes, only the central lower portion of the domain is shown, approximately corresponding to $1/8\le x/H\le7/8$ and $z/H\le0$ (length of scale bar is 2 cm). 
    }
    \label{fig:res1}
\end{figure}

\begin{figure}
    \centering
    \includegraphics[width=0.9\columnwidth, angle=0]{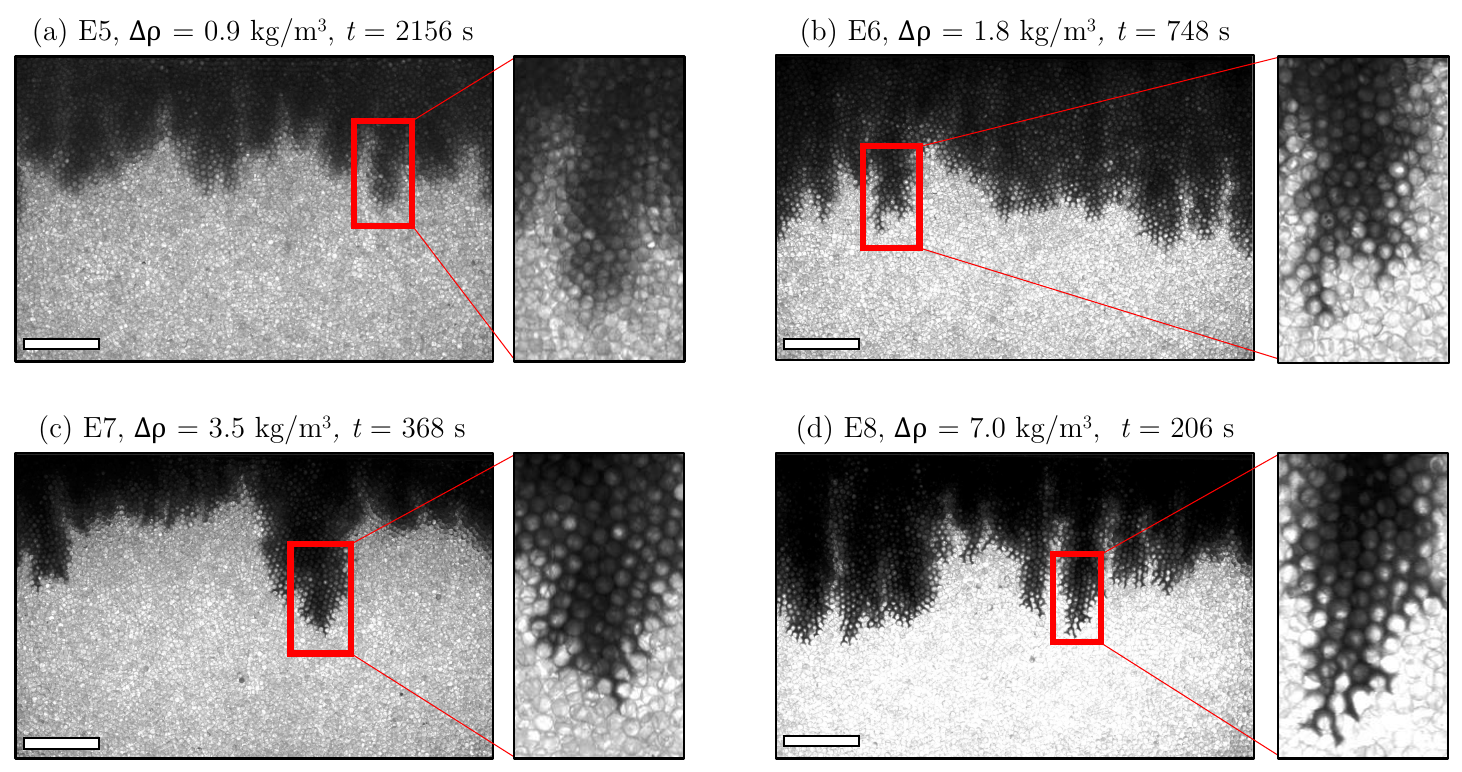}
    \caption{
    Flow evolution (raw images of laboratory experiments) for the same bead size, $d=1.75$~mm, and different density differences, $\Delta\rho$.
    The driving force is increased while the permeability remains unchanged.
    The flow structure changes remarkably from (a) to (d), due to the change in the size of the fingers relative to the beads.
    For visualisation purposes, only the central lower portion of the domain is shown, approximately corresponding to $1/8\le x/H\le7/8$ and $z/H\le0$ (length of scale bar is 2 cm). 
    }
    \label{fig:res2}
\end{figure}

Under certain flow conditions, a fluid flow through a porous medium may be considered as a Darcy flow.
In a Darcy flow, the length scale of the flow structures is much greater than the representative volume over which the quantities are averaged \citep{hewitt2020vigorous}.
This representative volume typically includes a number of solid particles and the interstitial fluid.
Darcy conditions are met when (i)~the flow is controlled by viscous forces at the pore scale $(\Rey\ll 1)$, and (ii)~the length scale of the convective flow is large compared to the pore size $(\pe\ll 1)$.
One can observe from table~\ref{tab:listex3} that only few experiments (E1-E3, E5) fall in the Darcy case.
A qualitative observation of this result is provided by looking at the raw images in figure~\ref{fig:res1} and ~\ref{fig:res2}.

In figure~\ref{fig:res1}, we consider the variation of the flow topology for the same driving force ($\Delta\rho\approx7$~kg/m$^3$) and different values of permeability (i.e., different $d$).
Both $\Rey$ and $\pe$ are sensitive to variations of the diameter of the beads.
However, $\Rey$ remains reasonably lower than unity for E4, E8, E12 and E16, whereas the Peclet number changes remarkably up to $O(10^{2})$.
This reflects the fact that the length scale of the convective flow is of the same order or smaller than the pore size. 
The transition is apparent when going from E4 in figure~\ref{fig:res1}(a), where the fingers width is at least a few diameters, to E12 in figure~\ref{fig:res1}(c), where one single plume penetrates and branches into the interstitial space.
In E16, shown in figure~\ref{fig:res1}(d), the permeability is considerably larger than in previous cases.
As a result, the driving force is extremely vigorous compared to the viscous drag, with the velocity $U$ approximately twice larger than in E12 (see table~\ref{tab:listex2}).
With $\pe\approx150$, the solute spreads quickly also in the direction perpendicular to the view, which makes the light intensity field blurry.

In figure~\ref{fig:res2}, we consider the opposite scenario, i.e., for the same permeability ($d=1.75$~mm) we vary the driving force (different $\Delta\rho$).
In this case, there is an increase in $\Rey$, which however remains considerably lower than 1.
The Peclet number increases of about one order of magnitude between E5 and E8, and we can appreciate this smooth transition from the intensity fields. 
The $\pe$ of E5 is just above unity and the width of the flow structures, shown in figure~\ref{fig:res2}(a), is about $5d-10d$.
Increasing the density difference makes the structures progressively smaller (E6, E7), and eventually the fingers propagate as the thin filamentary plumes (E8) shown in figure~\ref{fig:res2}(d).
Although the characteristic size of these plumes is comparable to the pore diameter, splitting in multiple branches is still observed.

\begin{figure}
    \centering
    \includegraphics{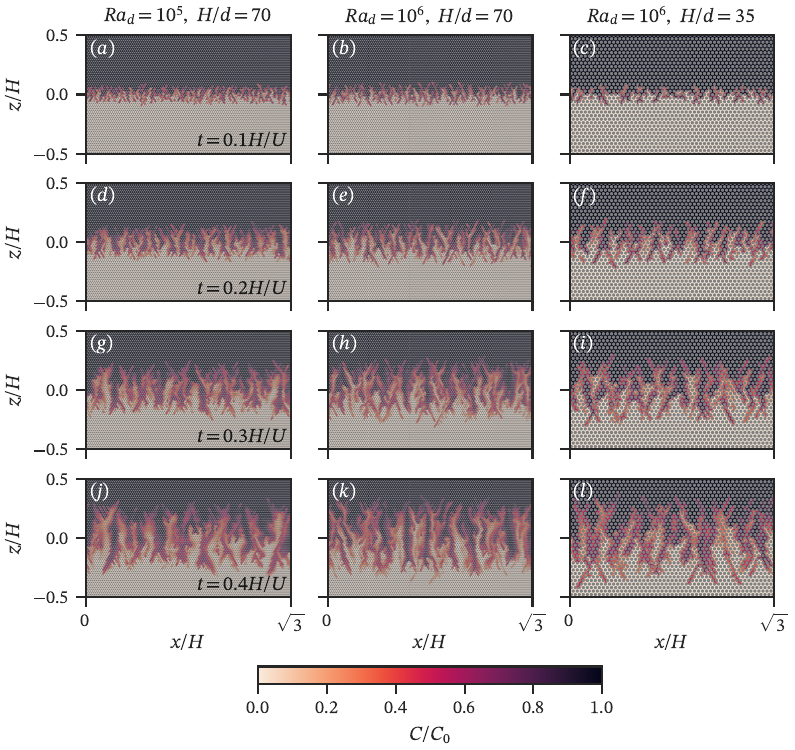}
    \caption{
        Snapshots of the concentration field from simulations S10, S12, and S6.
        In the first two simulations, $H/d=70$ is fixed, and only the Rayleigh number $Ra_d$ is varied from $10^5$ to $10^6$ (left to centre).
        In the third simulation (S6, right column), $Ra_d=10^6$ matches the centre column but $H/d$ is decreased to 35, so the obstacles are larger compared to the domain height.
        Each snapshot is presented at a given fraction of the timescale $H/U$, with time increasing top to bottom (as specified in the left column).
        Videos showing the evolution of the concentration field in selected simulations (S4, S6, S12) are available in the supplementary material (movie~2, movie~3 and movie~4, respectively).
    }
    \label{fig:sim_snaps}
\end{figure}

We have shown that the experimental results may present in some cases features that can be captured by a Darcy model when $Pe$ is low.
By contrast, the Peclet number for each numerical simulation satisfies $Pe\geq 5$, producing a flow that does not fulfil the Darcy assumptions outlined in Eq.~\eqref{eq:eqpere}.
Indeed, thin fingers penetrating the pore-space are always observed in the simulations, as shown in figure \ref{fig:sim_snaps} through a series of snapshots of the concentration field for various pore-scale Rayleigh numbers $Ra_d$.
In each case, the early-time snapshots highlight an initial instability on the scale of the beads, with thin plumes growing from the interface at $z=0$.
As time goes on, larger structures begin to develop in the mixing layer, with coherent fingers spanning the centre of the domain.
Even as these large-scale fingers grow, the dynamics at the tips of the fingers continues to consist of thin percolating structures, similar to what is observed in the experiments in figures \ref{fig:res2}(b)-(d).

Comparing the first two columns in figure~\ref{fig:sim_snaps} with each other, the main effect of changing $Ra_d$ is in the lateral diffusion of the concentration field.
The late-time concentration observed in panel (j) appears somewhat smeared out, with smoother gradients compared to the equivalent panel at higher $Ra_d$, panel (k).
The same Rayleigh number but a different geometry of the medium are considered in the central and right columns.
The minimum structure size is set by the pore-space, and the width of the fingers (relative to $H$) increases from $H/d=70$ (central column) to the $H/d=35$ (right column).
In all the simulations, the concentration field is strongly influenced by the hexagonal lattice structure of the medium, with fingers percolating aligned at an angle.
Although figure \ref{fig:sim_snaps} only features simulations with regular bead patterns, we shall show in the following sections that quantitative measures of the mixing are not significantly affected by the random obstacle perturbations shown in figure \ref{fig:pore_sim_schematic}.
Furthermore, when time is scaled with $H/U$ as in figure \ref{fig:sim_snaps}, the size of the mixing layer appears consistent across all three simulations, regardless of the control parameters $Ra_d$ and $H/d$.
We proceed to analyse this more quantitatively in the following section.

\section{Mixing length and concentration profiles}\label{sec:results2}
The mixing length is defined as the vertical extent of the region where solute concentration is non-uniform. 
Multiple ways to quantify the mixing length have been proposed \citep{cook2004mixing}, depending on whether a bulk-focused measure is taken or it tracks the spread of the fastest growing fingers \citep[e.g., $0.01\le \overline{C}/C_0 \le 0.99$, see][]{gopalakrishnan2017relative}. 
The latter method is used here to determine the mixing length obtained from the experiments, where a threshold value corresponding to $0.96$ is applied to the horizontally-averaged corrected light intensity as defined in~\eqref{eq:norm2}, $\overline{I}_2$.
The mixing length is then computed assuming that the flow is symmetric with respect to the domain centerline, and a time correction is also applied (see~\S\ref{sec:appa2}).

\begin{figure}
    \centering
    \includegraphics{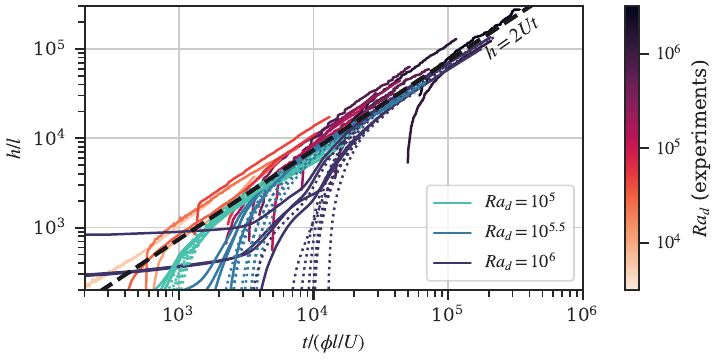}
    \caption{\label{fig:ml1}
    Dimensionless mixing length scaled with respect to $\ell$ for experiments (colorbar) and simulations (legend). 
    Simulations are shown for the regular pattern (solid lines) and the randomly perturbed pattern (dotted lines) as detailed in figure \ref{fig:pore_sim_schematic}.
    In the experiments, mixing length is computed assuming that the flow is symmetric with respect to the domain centerline, and a time correction is also applied (see~\S\ref{sec:appa2}).     
    Data is only plotted up to the time when the fingers reach the edge of the domain, so when $h=H$.
    Asymptotically, experiments and simulations follow the scaling $h=2Ut$ (dashed line).
    }
\end{figure}

The threshold-based definition discussed above would track the vertical extremes of the finger growth over time.
Alternatively, we can define a mixing length based on the mean concentration profile, and this is the approach we use for the simulations.
For this, we can assume that the mean (i.e, horizontally-averaged) profile takes a piecewise linear profile
\begin{equation}
    \frac{\overline{C}(z,t)}{C_0} =
    \begin{cases}
        0 & z\le-h/2 \\
        1/2 + z/h & |z| < h/2 \\
        1 & z\ge h/2
    \end{cases} . 
    \label{eq:pw}
\end{equation}
Here, the overbar denotes a horizontal average, $h$ is the mixing length, and the initial interface position is taken as $z=0$.
This assumed profile satisfies the integral relation
\begin{equation}
    h = \frac{6}{C_0^2}\int_{-\infty}^{\infty} \overline{C} (1 - \overline{C}) \,\mathrm{d}z ,
    \label{eq:mlnum}
\end{equation}
which we can use as a systematic definition of the mixing length $h$ in the simulations.
This integral definition provides a more useful statistical measure of the mixing layer than the threshold-based definition, but is impossible to compute with good accuracy from the experiments due to the considerable thickness of the cell and the opacity of the dye.

The dimensionless mixing length scaled with respect to $\ell$ is shown for experiments (colorbar) and simulations (legend) in figure~\ref{fig:ml1}. 
Simulations are presented for both the regular bead pattern (solid lines) and the perturbed bead pattern (dotted lines) as shown in figure~\ref{fig:pore_sim_schematic}.
Asymptotically, both experiments and simulations follow the scaling $h=2Ut$ (dashed line).
The buoyancy velocity $U$ represents the terminal velocity of a rising (falling) parcel of light (heavy) fluid surrounded by heavy (light) fluid, and it is achieved at the equilibrium between the driving force, represented by buoyancy, and the drag due to viscous forces within the medium.
This model is a simplified representation of the flow, and any diffusion effect or interaction with the flow structures is neglected.
However, it provides a good first order estimate for the growth of the mixing region. 
The observed behaviour gives a precise indication of the flow dynamics: after a starting phase in which the flow is influenced by the initial condition, the late-stage dynamics is controlled by convection, with the mixing layer growing at the characteristic buoyancy velocity $U$.

In three-dimensional porous systems, \citet{sardina2018buoyancy} showed that the mixing layer grows linearly for low values of porosity ($\phi=0.6$), while it follows the classical turbulent quadratic scaling when the porosity approaches 1.
This suggests that a linear growth of the mixing region is expected also for the values of porosity considered in this study.
However, the additional degree of freedom provided by the third spatial dimension may have an effect on the velocity at which the plumes spread: compared to the two-dimensional case, more pathways will be available to the individual fingers, which can spread more horizontally. 
As a result, we expect that during the convective regime the mixing length will still grow at a rate that is constant in time and lower compared to the two-dimensional case.
A uniform growth of the mixing layer is also observed in two-dimensional and three-dimensional Darcy flow at large Rayleigh-Darcy numbers \citep{Boffetta2020,borgnino2021dimensional}.
Also in this case, when comparing the evolution of two- and three-dimensional flows some differences emerge.
The growth rate of the mixing length is larger in two dimensions than in three dimensions \citep{borgnino2021dimensional}, and the transition between these regime occurs sharply \citep{boffetta2022dimensional}, as soon as the domain thickness is larger than the wavelength of the most unstable mode. 
The nature of this transition in pore-resolved flows has not been explored yet.
At the Darcy scale and at moderate Rayleigh-Darcy numbers, a super-linear scaling has been observed \citep{depaoli2019prf}, which has been interpreted has a finite size effect \citep{boffetta2022dimensional,depaoli2022experimental}, i.e., the time/space available for the fingers before touching the horizontal boundaries is insufficient to reach this asymptotic regime.

\begin{figure}
    \centering
    \includegraphics{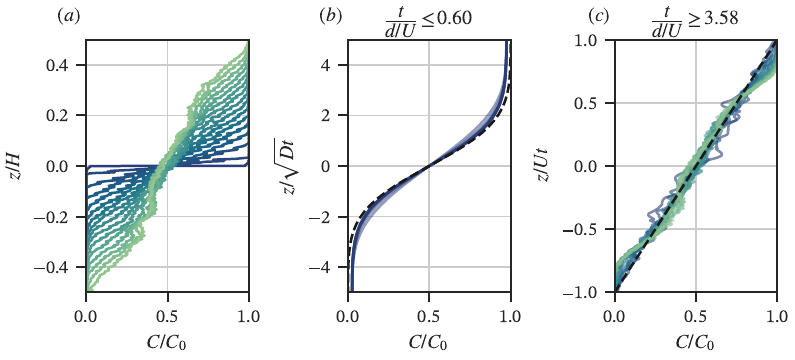}
    \caption{
    Vertical profiles of the horizontally-averaged concentration field $\overline{C}(z,t)$ for simulation S12: $(a)$~plotted over the domain height; $(b)$~plotted against the diffusive similarity variable; $(c)$~plotted against height rescaled using the buoyancy velocity.
    }
    \label{fig:sim_mean_conc}
\end{figure}

In the experiments, variations of solute concentration within the mixing layer cannot be inferred with good accuracy.
By contrast in the simulations, we have all the information to quantify how the solute is distributed across the mixing region.
Specifically, in figure \ref{fig:sim_mean_conc} we show the time evolution of the horizontally averaged concentration profile $\overline{C}(z,t)$ for a given simulation, in this case S12.
This averaging is only performed over the fluid fraction of the domain.
In figure \ref{fig:sim_mean_conc}(a), we present the mean concentration over the full height of the domain.
The profile is constant in the regions above and below the mixing layer, and transitions between 0 and $C_0$ in between, with this mixing layer expanding over time.
At very early times, before the emergence of the Rayleigh-Taylor instability, we expect the mean profile to develop diffusively.
This is confirmed by figure \ref{fig:sim_mean_conc}(b), where the early-time profiles of mean concentration are plotted against a rescaled spatial variable $z/\sqrt{Dt}$.
The profiles collapse, suggesting a self-similar development at this stage, and agree well with the analytic solution for a diffusing interface shown by the dashed black line.
The profiles do not perfectly tend to 0 and $C_0$ in this panel since the thin diffusive interface is contained within the region seeded with initial noise.
Once the Rayleigh-Taylor instability develops and saturates, the dynamics are controlled by a balance between the buoyancy driving and the drag provided by the porous medium.
We therefore expect the buoyancy velocity scale $U$ as defined in \eqref{eq:buoy_velocity} to play an important role in the spread of the solute.
Indeed, by plotting the mean concentration against the rescaled dimensionless coordinate $z/Ut$, we observe further a self-similar behaviour, with $\overline{C}$ remaining close to a linear profile within the mixing layer.
Since the result of figure~\ref{fig:sim_mean_conc}(c) is consistent with the assumption of \eqref{eq:pw}, we are confident that the resulting mixing length expression \eqref{eq:mlnum} is reliable for our simulations.

\section{Flow structure and wavenumber}\label{sec:results3}

The flow structure is first discussed qualitatively with the aid of experimental measurements, and then quantitatively using the numerical results.

\begin{figure}
    \centering
    \includegraphics[width=0.98\columnwidth, angle=0]{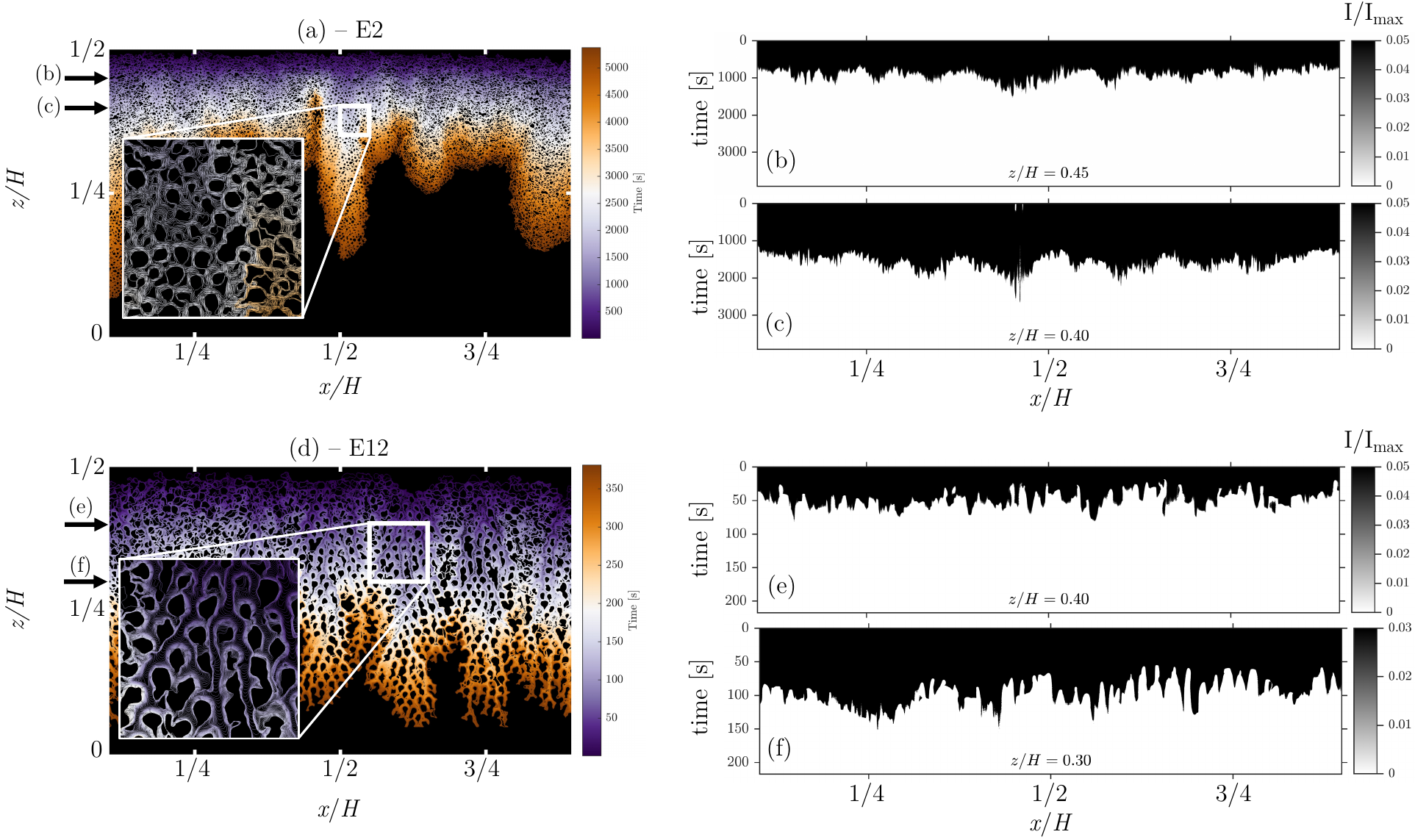}
    \caption{
Interface and flow structure evolution for two different experiments. 
The instantaneous fluid-fluid interface is reported from early (purple) to late (brown) times (see also Movie 1 in the supplementary material).
For visualisation purposes, only a portion of the entire domain width is shown.
Experiment E2 is shown in (a), where the inset corresponds to a squared region of side $H/20$.
Panels~(c) and (d) report the space-time intensity maps taken at $z/H=-0.05$ and $z/H=-0.10$, respectively, as indicated with the black arrows in panel~(a), obtained from the local values of raw light intensity ($I$) at that height $z$ normalised by the maximum value within the picture ($I_\text{max}$).
The time-dependent interface evolution for experiment E12 is reported in panel~(d), where the inset corresponds to squared region of side $H/10$.
Here the width of the flow structures is comparable with the interstitial space, i.e. with the diameter of the beads. 
Space-time maps of normalised intensity taken at different heights (panels~e,f) reveal more clearly that in this case the flow has a behaviour that is very different from E2, with thin fingers penetrating individually into the unmixed region.
    }
    \label{fig:wn1}
\end{figure}

In figure~\ref{fig:wn1} we consider the interface evolution for two different experiments, namely E2 and E12. 
In figure~\ref{fig:wn1}(a), the evolution of the fluid-fluid interface is reported from early (blue) to late (brown) times.
The interface is computed from the normalised intensity fields $I_1$, as discussed in \S\ref{sec:meth3a}.
The flow around the beads is recorded at a high resolution, and the position of the interface is reconstructed.
An example of the behaviour of the interface at the scale of the pores is reported in the inset of figure~\ref{fig:wn1}(a), where a squared region with side $H/20$ is magnified.
From this qualitative view of the evolution of the interface, we observe that the structures of the convective flow are much larger than the beads diameter. 
To more quantitatively evaluate the interface shape at different heights, we consider the space-time maps in  figures~\ref{fig:wn1}(b) and (c), taken at $z/H=-0.05$ and $z/H=-0.10$, respectively, as indicated with the black arrows in figure~\ref{fig:wn1}(a).
These maps are obtained from the local values of raw light intensity ($I$) at that height $z$ normalised by the maximum value within the picture ($I_\text{max}$).
We observe that, at both heights, the characteristic wavelength of the interface is larger than the beads size, which is a feature typically observed in Darcy-type flows.
In addition, we also observe that the lower the measurement location, the later the interface appears, and the larger the amplitude of the interfacial oscillations. 
Similarly, the time-dependent interface evolution for experiment E12 is reported in figure~\ref{fig:wn1}(d).
In this case, the flow corresponds to a high-$\pe$ flow, and the wavelength of the interfacial structures are comparable in size with the interstitial space, i.e. with the diameter of the beads. 
Space-time maps of normalised intensity taken at different heights (figures~\ref{fig:wn1}e,f) reveal more clearly that in this case the flow has a behaviour that is very different from the previous one, with thin fingers penetrating into the unmixed region.
This analysis, although qualitative, provides information about the flow structure in different conditions, from a dissipation-controlled flow (E2) to a buoyancy-controlled flow (E12).

\begin{figure}
    \centering
    \includegraphics{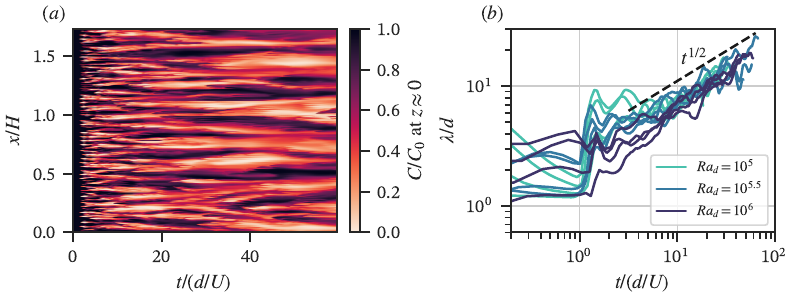}
    \caption{(a) Space-time plot of the concentration profile $C(x,t)$ just above the initial interface height $z=0$ from simulation S12. (b) Time series of the mid-height finger wavelength as calculated using \eqref{eq:wavelength_fft} for each simulation with a regular bead pattern.
    }
    \label{fig:wavelengths}
\end{figure}

We now turn to the simulations to provide some quantitative results on the wavelength of the finger structures.
Since we can only capture the lower edge of the interface in the experiments, it is not possible to investigate the evolving dynamics of the fingers at the centre of the mixing layer.
By contrast, in the simulations we have the full details of the flow field, and can make more quantitative statements about the time evolution of the finger structures.
We track the evolution of the finger width by considering the concentration profile in horizontal cuts near the initial interface position $z=0$.
In the cases where we position the solid beads in a regular hexagonal packing, we can take a horizontal cut through the domain that only contains the liquid phase.
This is important for analysis of the finger width, for which we will take advantage of the horizontal periodicity and use Fourier transforms.
Specifically, we take the Fourier transform of the concentration field in the $x$-direction at a fixed height $z_m$ to obtain $\hat{C}(k,t)$, and then define the mean horizontal wavelength of the plumes as
\begin{equation}
    \lambda(t) = \frac{\int_0^\infty k^{-1} \left|\hat{C}(k,t)\right|^2 \,\mathrm{d}k}{\int_0^\infty \left|\hat{C}(k,t)\right|^2 \,\mathrm{d}k} \label{eq:wavelength_fft} .
\end{equation}
As shown in figure \ref{fig:wavelengths}, the finger structures at the centre of the domain exhibit a coarsening behaviour, with the mean wavelength increasing over time.
This coarsening is consistent across all the simulations at varying $Ra_d$ and $H/d$.
After initial transient phase related to the onset of the Rayleigh-Taylor instability, the wavelength exhibits a power-law scaling of $t^{1/2}$.
Such a scaling is often associated with a diffusive process, although in figure~\ref{fig:wavelengths} the collapse is observed in terms of the advective scale $U$ rather than the diffusivity $D$.
This result of a scaling close to $t^{1/2}$ is in fair agreement with the Darcy simulations of \citet{depaoli2019universal}, where the mean wavenumber $\kappa\sim1/\lambda$ exhibits a $t^{-0.6}$ scaling during the growth of the mixing layer.
\cite{Boffetta2020} also observe $t^{1/2}$ scaling for the horizontal wavenumber in three-dimensional Darcy simulations of Rayleigh-Taylor instability, finding a collapse with $\lambda \sim \sqrt{Dt}$.
This diffusive behaviour is attributed to nonlinear processes, including plume merging and coarsening, and it is not a simple direct consequence of molecular diffusion.
This contrasts to our pore-scale simulations, where the rate of coarsening is independent of the molecular diffusivity.
As the mixing layer grows in our simulations, and the fingers coarsen to scales larger than a few multiples of the pore scale, the Darcy assumption seems to become more relevant, as would be expected.

\section{Mean scalar dissipation}\label{sec:results4}
\begin{figure}
    \centering
    \includegraphics{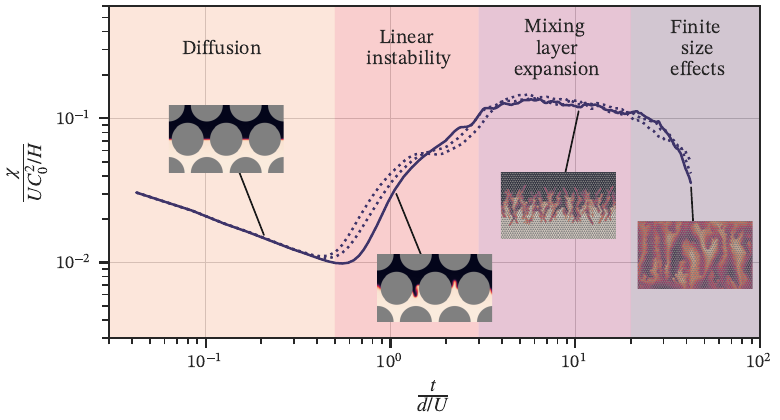}
    \caption{
    Evolution of the mean scalar dissipation ($\chi$) over time ($t$) for simulation S6. 
    The time is made dimensionless with $(d/U)$. 
    Three different simulations are shown, corresponding to the parameters as reported in table~\ref{tab:listex3} and different arrangements of the obstacles (regular and perturbed patterns, corresponding to solid and dotted lines respectively).
    A few concentration fields in different phases of the process are also shown (regular pattern).
    }
    \label{fig:dissipation_rate}
\end{figure}

Numerical simulations allow accurate local measurements of concentration gradients. 
They are used here to infer the local mixing state of the system via the mean scalar dissipation,
\begin{equation}
\label{eq:chi}\\ 
 \chi = D \langle |\nabla C|^2 \rangle,
\end{equation}
where $ \langle \cdot \rangle$ stands for volume average over the fluid volume.
The evolution observed for $\chi$ is similar across all the considered simulations, and it is illustrated in figure~\ref{fig:dissipation_rate} for simulation S6.
We observe that four main regimes can be identified here: (i) an initial diffusive regime, followed by (ii) a linear instability, (iii) a convection-driven regime, and (iv) a final shutdown phase.
In this section, we will discuss in detail these phases of the mixing process.

The mean scalar dissipation can be related to other global quantities of the flow \citep{hidalgo2012scaling,depaoli2023convective}, such as the dissolution rate (in case of systems permeable to solute at the boundaries) or the volume-averaged squared concentration, $\langle C^2\rangle$.
Given our no flux boundary conditions for concentration, the relation
\begin{equation}
        \partial_t \langle C^2\rangle = -2 \chi, \label{eq:conserv}
\end{equation}
holds and will be used in the following to describe the evolution of the dissipation rate.

\begin{figure}
    \centering
    \includegraphics{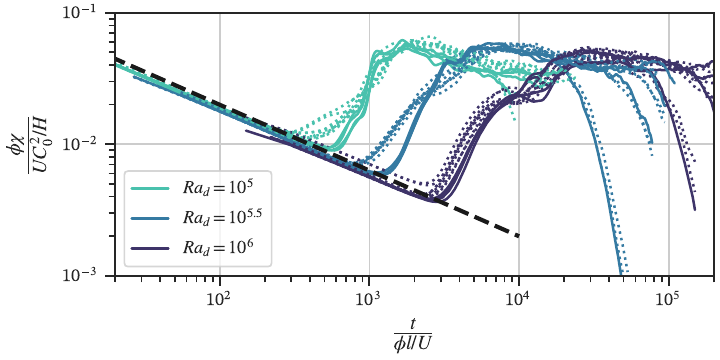}
    \caption{
    Dimensionless mean scalar dissipation rate over dimensionless time obtained from the numerical simulations (with regular and perturbed bead patterns, corresponding to solid and dotted lines respectively).
    Black dashed line indicates the diffusive evolution derived in equation~\eqref{eq:diff4}. 
    }
    \label{fig:dr1}
\end{figure}

We now consider the volume-averaged scalar dissipation rate over time for all simulations, reported in figure~\ref{fig:dr1}.
A natural length scale to be used to analyse the results during the diffusive regime is $\ell$, defined in \eqref{eq:eq34}, and therefore time is scaled with $\phi\ell/U$.
In this frame, all simulations nicely collapse onto the same curve in the initial diffusive phase, and we provide in the following an analytical description of the mixing process. 

The fluid is initially motionless, with a step-like concentration field.
As a result, one can neglect any velocity contribution and assume that the concentration field is uniform in horizontal direction. 
It follows that the initial development, $t/(\phi\ell/U)<3\times 10^{3}$, is described by a purely diffusive 1-D solution, where
\begin{equation}
    C = \frac{C_0}{2}\left[1 +\mathrm{erf} \left( \frac{z}{2 \sqrt{D t}}\right)\right] .
    \label{eq:diff1}
\end{equation}
From this expression, we can derive the scalar dissipation rate by
\begin{equation}
    \chi =  D \langle |\nabla C|^2 \rangle = \frac{ D}{H} \int_{-H/2}^{H/2} |\partial_{z} C|^2 \mathrm{d}z 
    \approx \frac{ D}{H} \int_{-\infty}^{\infty} |\partial_{z} C|^2 \mathrm{d}z =
    \sqrt{\frac{ D}{8 \pi t}} \frac{C_0^{2}}{H}.\label{eq:diffusive_gradient2}
\end{equation}
This expression made dimensionless with $UC_0^{2}/(\phi H)$ can be expressed as a function of $t/(\phi\ell/U)$ as:
\begin{equation}
   \frac{\phi\chi}{UC_0^{2}/H} =\frac{1}{\sqrt{8\pi}}  \left(\frac{t}{\phi\ell/U}\right)^{-1/2}.
   \label{eq:diff4}
\end{equation}
The corresponding behaviour is shown in figure~\ref{fig:dr1} (black dashed line).
It is in agreement with the numerical findings since all simulations, regardless of their $Ra_{d}$ and $H/d$, nicely collapse onto the analytical prediction.
The same solution obtained in equation~\eqref{eq:diffusive_gradient2} can be derived using equation~\eqref{eq:conserv} with the diffusive solution~\eqref{eq:diff1}, which gives:
\begin{equation}
\chi = \sqrt{\frac{ D}{8 \pi t}} \frac{C_0^{2}}{H}g(\xi),
\quad\text{with}\quad
g(\xi) = \text{erf}(\xi)-\text{erf}\left(\frac{\xi}{\sqrt{2}}\right)\sqrt{2e^{-\xi^2}}
\label{eq:diffusive_means}
\end{equation}
and $\xi=H/(2\sqrt{Dt})$. 
Note that during the diffusive regime $\xi\gg1$ and $g(\xi)\approx1$, consistent with \eqref{eq:diffusive_gradient2}.

The system leaves this diffusive behaviour as soon as convective instabilities develop, i.e., when finger-like structures form. 
These structures stretch the interface, providing a larger area for diffusion to act over and thus promoting mixing (with a corresponding increase in $\chi$).
The time at which the instabilities become significant depends both on the magnitude of the initial perturbation applied to the concentration field, and on the extension of the region at which this perturbation is applied.

At later times, buoyancy-driven fingers formed at the initial fluid-fluid interface grow towards the horizontal walls. 
This process is on one hand controlled by convection, which drives the fingers elongation and the corresponding increase of their interfacial extension, and on the other hand by diffusion, which reduces the concentration gradient across the fingers interface in time \citep{gopalakrishnan2017relative}.
In addition, the presence of solid obstacles makes the fingers more prone to split, possibly increasing further the fluid mixing.

\begin{figure}
    \centering   
    \includegraphics{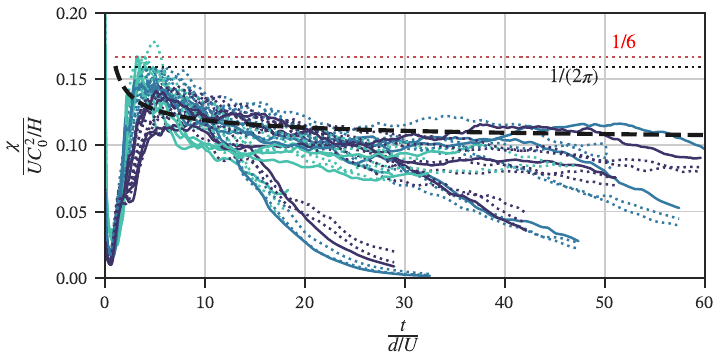}
    \caption{
    Dimensionless mean scalar dissipation rate over dimensionless time obtained from the numerical simulations (with regular and perturbed bead patterns, corresponding to solid and dotted lines respectively, colors as in Fig.~\ref{fig:dr1}).
    The black dotted line, corresponding to a dimensionless value of $1/2\pi$ from \eqref{eq:diff3}, refers to the maximum mean dissipation rate in the convective regime.
    The alternative estimate of $1/6$ derived in \eqref{eq:conv_e} is also shown as a red dotted line.
    The thicker black dashed line denotes the time-dependent decreasing mean dissipation, as predicted by \eqref{eq:int5}, in the same regime.
    The fitting parameter $\beta=0.75$ is chosen such that the estimate tends to 0.1 as $t\rightarrow\infty$.
    }
    \label{fig:dr2}
\end{figure}

We provide a first estimate for the maximum dissipation rate in the convective regime by considering that dissipation mainly takes place within the mixing layer, where local non-zero concentration gradients exist.
Outside the mixing layer the fluid is nearly homogeneous in concentration.
The volume-averaged dissipation rate can therefore be approximated as
\begin{equation}
    \chi =  D \langle |\nabla C |^2 \rangle \approx  D \frac{h}{H} \langle |\nabla C|^2 \rangle_{ML} ,
\end{equation}
where $\langle \cdot \rangle_{ML}$ denotes an average value across the mixing layer.
By assuming that the convective fingers stretch the interface around the mixing layer, we can approximate the mean scalar gradient in the mixing layer as the gradient of \eqref{eq:diff1} at the interface, thus as
\begin{equation}
    |\nabla C| \approx \frac{C_0}{2\sqrt{\pi  D t}} . \label{eq:scalar_gradient}
\end{equation}
Combining these approximations with the result of \S\ref{sec:results2} for the growth of the mixing region $(h \approx 2 U t)$, we arrive at the estimate
\begin{equation}
    \chi \approx  D \frac{2 U t}{H} \frac{C_0^{2}}{4\pi D t} = \frac{1}{2\pi} \frac{UC_0^{2}}{H} .
    \label{eq:diff3}
\end{equation}
This expression (black dotted line) proves to be an overestimate in figure~\ref{fig:dr2}.
The overestimation is expected since, in the frame of the interface, the approximation given by \eqref{eq:scalar_gradient} is the \emph{maximum} gradient rather than the average over a certain scale.

Similarly to what is observed during the diffusive regime, the dissipation rate can also be estimated here from the bulk squared concentration.
In the following argument, we take the exact relation \eqref{eq:conserv} and assume that $d_t\langle C^2\rangle \approx d_t \langle \overline{C}^2\rangle$.
This implies that the leading order effect of dissipation is on mixing the mean concentration $\overline{C}$, but does \emph{not} mean that dominant contribution to the dissipation is \emph{from} the mean profile.
We recall from figure~\ref{fig:sim_mean_conc}(c) that, during the convective regime, the mean concentration field is linear within the mixing region.
Then we can approximate the horizontally-averaged concentration as:
\begin{equation}
\frac{\overline{C}(z,t)}{C_0} = \begin{cases}
  0  & \text{for} -H/2\le z\le -h/2\\
  1/2+z/(2Ut) & \text{for } |z|< h/2\\
  1 &  \text{for } h/2\le z\le H/2
\end{cases},
\label{eq:avg_conc_lin}
\end{equation}
which is nothing but \eqref{eq:pw} with $h=2Ut$.
Using equation~\eqref{eq:conserv} with approximation~\eqref{eq:avg_conc_lin}, we estimate the mean scalar dissipation as
\begin{equation}
    \chi \approx -\frac{1}{2} \frac{\mathrm{d}}{\mathrm{d}t}\left(\int_{-H/2}^{-h(t)/2}\overline{C}^2 \mathrm{d}z+\int_{-h(t)/2}^{h(t)/2}\overline{C}^2 \mathrm{d}z+\int_{h(t)/2}^{H/2}\overline{C}^2 \mathrm{d}z\right) = \frac{1}{6} \frac{UC_0^{2}}{H} .
    \label{eq:conv_e}
\end{equation}
This result, shown as a red dotted line in figure~\ref{fig:dr2}, is analogous to the expression derived in equation~\eqref{eq:diff3}.
The obtained numerical coefficient is also similar.
As mentioned above, this estimate does not take into account local fluctuations of concentration, likely responsible of the decrease observed for the scalar dissipation, and to that aim higher order statistics should be considered.

When plotted on a linear scale, as in figure~\ref{fig:dr2}, it becomes clear that the dissipation rate decreases with time. 
The mechanism that controls the dissipation rate in this phase is possibly due to several interplaying phenomena.
While the mixing region grows at a constant speed ($h \approx 2 U t$), the concentration gradients across the fingers' interface reduce, owing to diffusion. 
In addition, the number of fingers varies in time, as well as the extension of the interfacial region across which the fluids mix.
We propose here a model that takes into account these features of the flow and explains the decreasing behaviour of $\chi$ in the convective phase.

\begin{figure}
    \centering
    \includegraphics[width=0.88\columnwidth]{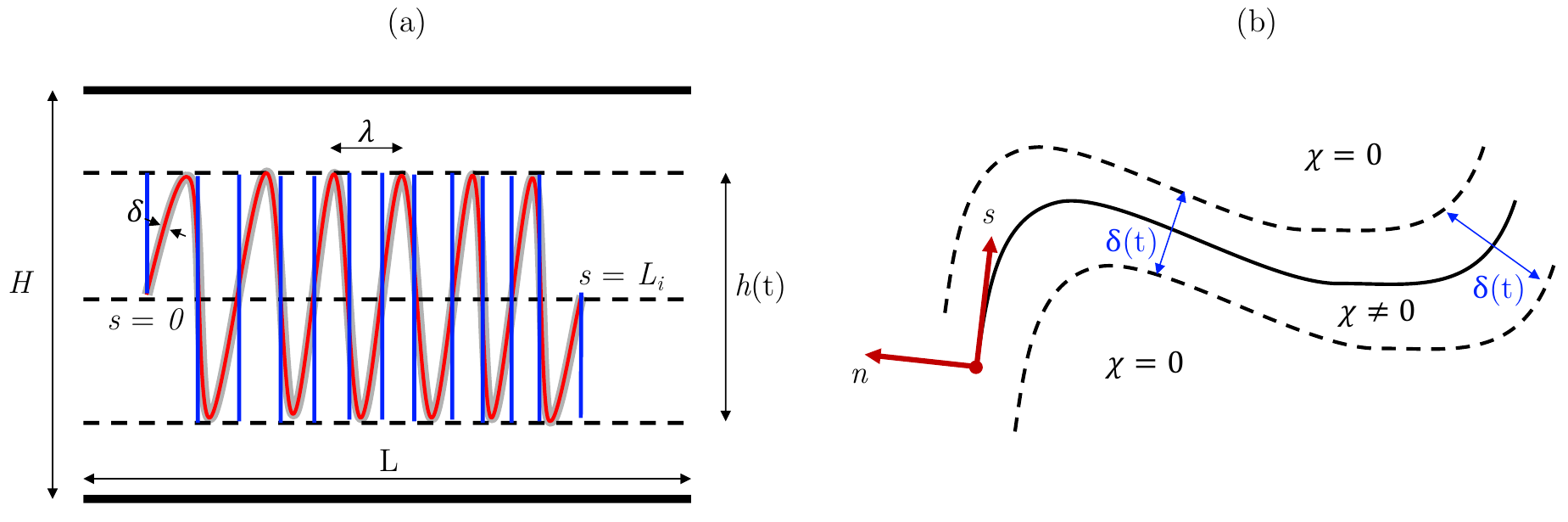}
    \caption{
    Model for the evolution of the interface in the convective phase.
    (a)~The entire domain (dimensions $L\times H$) is sketched for a time $t$ within the convective regime, when fingers have already developed. The average width of the fingers is $\lambda$.
    At this time, the extension of the mixing region is $h(t)$, and the interface between the fluids (red line) can be approximated by a segmented line (blue).
    (b)~Detail of the interfacial region. 
    We assume the interface (black solid line) has a finite and uniform thickness $\delta(t)$, and the mixing occurs within this region.
    A coordinate system is defined such that $s$ is tangential to the interface and $n$ perpendicular to it.
    }
    \label{fig:interfmodel}
\end{figure}

A schematic model for the fluid-fluid interface is proposed figure~\ref{fig:interfmodel}(a).
For any instant $t$ in the convective regime, fingers are considered to have all the same length, which matches the values of the mixing length $h(t)$. 
In addition, we also consider the fingers to have all the same width, $\lambda$. 
We approximate the fingers shape to be straight, and as a result the length of the interface $L_{i}$ to be equal to the segmented blue line in figure~\ref{fig:interfmodel}(a), which is given by:
\begin{equation}
    L_{i}(t) =L+2\frac{L}{\lambda(t)}h(t).
    \label{eq:int1}
\end{equation}
We introduce in figure~\ref{fig:interfmodel}(b) a coordinate system defined such that $s$ is tangential to the interface and $n$ perpendicular to it.
As a result, assuming an interface with uniform thickness $\delta(t)$, the mean scalar dissipation can be computed by integrating the $|\partial_{n} C|^{2}$ across the thickness $\delta$ and along the interface length $L_{i}$ to obtain
\begin{equation}
    \chi =  D \langle |\nabla C |^2 \rangle = \frac{DL_{i}}{HL}\int_{-\delta/2}^{+\delta/2}|\partial_{n}C|^{2}\diff n \approx \frac{D L_i}{HL} \frac{C_0^2 \delta}{4\pi Dt}.
    \label{eq:int2}
\end{equation}
Here, we assumed that the concentration gradient across the interface evolves diffusively using \eqref{eq:scalar_gradient}, and took $\delta/(2\sqrt{Dt})\ll 1$. 
Motivated by the observations of coarsening fingers in figure \ref{fig:wavelengths}(b), we assume a diffusive growth for $\delta$ with an effective diffusivity of $dU$, namely:
\begin{equation}
    \delta = \beta \sqrt{dU t}, 
    \label{eq:int4}
\end{equation}
with $\beta$ being a fitting parameter. 
Note that the effective diffusivity $dU$ determined here coincides with the effective longitudinal dispersion usually considered for bead packs \citep{tsinober2023numerical,liang2018effect}.
We now substitute the expressions obtained in \eqref{eq:int1} and \eqref{eq:int4} into the definition \eqref{eq:int2}, together with the evolution of the mixing length $h=2Ut$ and the wavelength measured from the simulations $\lambda/d=\alpha\sqrt{t/(d/U)}$, as in figure~\ref{fig:wavelengths}.
Finally, we obtain an expression for $\chi(t)$ in the convective regime:
\begin{equation}
    \chi=  \frac{C_0^{2}U}{H}\frac{\beta}{\alpha\pi}\left[1+\frac{\alpha}{4}\left(\frac{t}{d/U}\right)^{-1/2}\right],
    \label{eq:int5}
\end{equation}
in which $\alpha=2.39$ is obtained from the numerical results of figure \ref{fig:wavelengths} and $\beta=0.75$ is obtained as a fitting parameter for the data of $\chi(t)$.

Comparison of the analytical prediction \eqref{eq:int5} (thick black dashed line) against the numerical results of mean scalar dissipation is shown in figure~\ref{fig:dr2}.
This solution captures more accurately the evolution of the mixing process, and it is also quantitatively in agreement with the maximum dissipation \eqref{eq:diff3} (thin black dashed line) at early times.
This result suggests also that mixing process during the convective regime, unlike the diffusive regime, is controlled by the characteristic pore-scale of the domain, $d$. 
The solution is indeed independent of the domain size, $H/d$, and the driving force, $Ra_{d}$.

We observe that the behaviour of $\chi$ shown in figure~\ref{fig:dr2} is remarkably dissimilar compared to the corresponding Darcy simulations \citep{depaoli2019universal}, where it is observed to increase with time up to impingement of the fingers on the horizontal boundaries.
\citet{hidalgo2015dissolution} observed that in presence of fluids characterized by non-monotonic density-concentration profiles, which leads to a considerably different flow topology, the scalar dissipation remains constant during the convective regime.
We speculate that the pore-induced dispersion effects \citep{dentz2018mechanisms} are likely responsible for these differences.

\begin{figure}
    \centering 
    \includegraphics{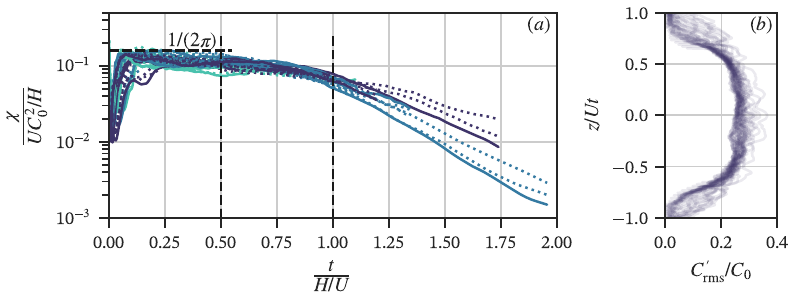}
    \caption{
    (a) Dimensionless mean scalar dissipation rate against dimensionless time obtained from numerical simulations (with regular and perturbed bead patterns, corresponding to solid and dotted lines respectively, colors as in Fig.~\ref{fig:dr1}).  
    The horizontal dashed line, corresponding to a dimensionless value of $1/2\pi$, equation~\eqref{eq:diff3}, refers to the maximum mean dissipation rate in the convective regime.
    Vertical dashed lines approximately correspond to the time required for the fingers to reach the horizontal walls of the domain, $t/(H/U)=1/2$, and for the core of the domain to be influenced by the presence of the walls, $t/(H/U)=1$.
    (b) Vertical profiles of the root-mean-squared concentration from simulation S12, highlighting the uniform region of concentration variance within the region $|z|<Ut/2$. Instantaneous profiles are plotted for $t/(U/d) \geq 5$, each with an opacity of $0.1$.
    }
    \label{fig:dr3}
\end{figure}

During the convective regime, the fingers grow under the action of buoyancy and eventually touch the horizontal boundaries of the domain.
This event occurs approximately at ${t=H/(2U)}$, given the growth rate of the mixing region, and marks a first observable reduction of the scalar dissipation. 
It is apparent that in this regime, referred here to as a shutdown regime, the relevant flow length scale is $H$. 
Therefore we report in figure~\ref{fig:dr3}(a) the evolution of $\chi$ as a function of $t/(H/U)$, and we observe that all curves nicely collapse in the late stage of the flow evolution, i.e. for $0.25\le t/(H/U)\le1$.

After the fingers impinge on the horizontal walls, the concentration field in the near-wall regions begins to be progressively more homogeneous.
This reflects the reduction of the local concentration gradients, and therefore of the mean scalar dissipation.
Mixing is still ongoing, however, in the central portion of the domain, where the information that the walls are present has not reached yet.
This can be seen in figure \ref{fig:dr3}(b), where we plot the root-mean-squared concentration profiles for a typical simulation.
With height rescaled by $Ut$, we observe a `core' region of uniform scalar variance persists between $-Ut/2 \leq z \leq Ut/2$.
The shape of this profile persists even after the fingers reach the domain boundaries.
Approximately at time $t=H/U$, the core of the flow is affected as well, and the entire domain becomes more homogeneous in solute concentration, i.e., the local concentration gradients are small and the mean scalar dissipation drops further.
The overall dynamics is controlled in this case by the domain height, however geometry and buoyancy still play a role in determining the reduction rate of the scalar dissipation.

\section{Summary, conclusions and outlook}\label{sec:concl}

We have studied the process of convective dissolution in porous media.
We considered a Rayleigh-Taylor instability, consisting of two miscible fluid layers of different density placed in an unstable configuration, with the heavy fluid on top of the lighter one. 
The flow is unstable due to the presence of a solute, which induces the density differences driving the mixing process.
The porous medium consists of a confined, homogeneous and isotropic bead pack, with the solid obstacles being impermeable to fluid and solute.
We investigated the flow at the scale of the pores using experimental measurements, numerical simulations and physical models.
Simulations employ finite differences coupled to the immersed boundary method, while experiments are performed with transparent beads and fully miscible fluids. 
Experiments and simulations have been specifically designed to mimic the same flow conditions, namely linear dependency of fluid density with solute concentration, high Schmidt numbers ($\sch=O(10^2)$) and the same values of porosity.

The evolution of the flow is quantified by the mixing length $h$, which represents the vertical extension of the mixing region.
We demonstrate via experiments and simulations that the system is characterized by a linear scaling $h=2Ut$, and the growth of the mixing region is controlled by the buoyancy velocity $U$.
This velocity is achieved at the equilibrium between driving forces (density contrast between the fluids, $\Delta\rho$) and viscous dissipation (drag through the medium).
The solute evolution presents a self-similar behaviour during the initial diffusive phase and during the following convective regime.
We demonstrate this self-similarity of the flow by inspection of the horizontally-averaged concentration profiles.
The flow structures are analysed using the mean wavelength at the centreline ($\lambda$), and also in this case a scaling holds ($\lambda\sim\sqrt{Ut}$).
Finally, we analyse the mixing dynamics of the system, which is quantified via the mean scalar dissipation rate, $\chi$, and three flow regimes are observed.
The flow is initially controlled by diffusion $(\chi\sim\sqrt{D/t})$, which produces solute mixing across the initial horizontal interface.
When the interfacial diffusive layer is sufficiently thick and it eventually becomes unstable, finger-like structures form and drive the flow into a convection-dominated phase.
In this case, fluid mixing is controlled by diffusion (predominantly at the sides of the fingers) and by buoyancy-driven convection (which makes the finger grow vertically as $h=2Ut$).
After an initial growth at the end of which the dissipation rate attains a maximum value (which we predict), a reduction of the mean dissipation rate is observed as a result of the competing effect of merging of the fingers (negative contribution) that dominates over the growth of the mixing region (positive contribution).
To describe this behaviour, we propose a model of mixing that relies on the diffusion process across the interface of the fingers. 
Finally, when the fingers grow sufficiently to touch the horizontal boundaries of the domain, the scalar dissipation reduces dramatically (shutdown regime), due to the absence of fresh fluid contributing to mixing.
We further elucidated the physics of the observed phenomena with the aid of simple physical models, and we demonstrated that each regime is controlled by a different length scale, namely the scale of diffusion, the characteristic length of the pores, or the domain height.

In this study we have focused on modelling the mixing dynamics and the flow structure of porous media convection.
We used simulations performed in two-dimensional arrays of circular objects and experiments consisting of thin three-dimensional packs of spherical beads (Hele-Shaw type geometry).
In a porous Rayleigh-Taylor system, at the Darcy scale, the mixing region is observed to grow faster in two-dimensional domains than in three-dimensional ones \citep{borgnino2021dimensional}. 
Unlike in the turbulent case, the transition between these regimes occurs sharply when the thickness of the domain exceeds the wavelength of the most unstable mode \citep{boffetta2022dimensional}.
The nature of this transition in pore-resolved flows has not been explored yet, and in the future it will be of interest to extend the present study to simulations in three dimensions, as well as two-dimensional experiments \citep{de2022two}, to allow for a one-to-one comparison of these findings and to investigate the effect of the dimensionality of the flow on the evolution of a pore-resolved system.
At the pore scale, three-dimensionality provides more pathways for fingers to percolate through, and the interfacial area of three-dimensional fingers will be greater than for 2-D.
Such effects may potentially allow for greater dispersion, and quantifying the associated impact of the solute on the larger-scale spread will be a key focus of future simulations.

Present findings are relative to domains having a dimension up to few hundred pores.
In contrast, geophysical applications involve formations that may be orders of magnitude larger \citep{huppert2014fluid}, and modelling of the entire dissolution process may be required \citep{wang2022analysis,szulczewski2013carbon}.
For instance, in the case of carbon sequestration it is desired to determine the time required to dissolved a prescribed fraction of the injected fluid volume, or to estimate the horizontal spread of the current of CO$_2$ \citep{macminn2012spreading,depaoli2021influence}.
Performing pore-scale studies of such flows is beyond the present capabilities and Darcy simulations including the effect of dispersion are required \citep{dentz2023mixing}.
In this context, one remarkable finding of the present study is the behaviour of the mean scalar dissipation during the convection-dominated regime, which is observed to decrease in time in pore-resolved flows (see figure~\ref{fig:dr2}) while it grows in Darcy flows \citep{depaoli2019universal}.
A different behaviour is observed by \citet{hidalgo2015dissolution} in the presence of fluids characterized by non-monotonic density-concentration profiles.
In their Darcy simulations, this property of the fluids leads to a remarkably different flow topology with a nearly flat interface between the two fluid layers, and the scalar dissipation remains constant during the convective regime.
We speculate that the pore-induced dispersion effects \citep{dentz2018mechanisms}, not captured by Darcy simulations, are likely responsible for these differences and should be accounted by suitable dispersion models developed from pore-resolved data.

The results presented are relevant to homogeneous and isotropic porous media at the porosity $\phi=0.37$.
Geological formations are generally characterized by a more complex structure, typically heterogeneous at multiple scales \citep{hidalgo2022advective}, anisotropic \citep{ennis2005onset} and with porosity values that can be as low as 0.05 \citep{bickle2017rapid}. 
Exploring the effect of such irregularities of the medium on the flow structure would represent an important extension of this study to be considered for future works.

\backsection[Supplementary movies]{Supplementary movies are available.}
\backsection[Data availability statement]{
The data presented in this work are openly available at \citet{dataset}.}
\backsection[Funding]{
This project has received funding from the European Union's Horizon Europe research and innovation programme under the Marie Sklodowska-Curie grant MEDIA agreement No.~101062123.
The work of CJH was funded by the Max Planck Center for Complex Fluid Dynamics.
We acknowledge PRACE for awarding us access to MareNostrum4 at the Barcelona Supercomputing Center (BSC), Spain and IRENE at Très Grand Centre de Calcul (TGCC) du CEA, France (project 2021250115).
This work was also carried out on the Dutch national e-infrastructure with the support of SURF Cooperative.
This research was funded in part by the Austrian Science Fund (FWF) [Grant J-4612].}

\backsection[Declaration of interests]{The authors report no conflict of interest.}
\backsection[Author ORCID]{
\\Marco De Paoli, \href{https://orcid.org/0000-0002-4709-4185}{https://orcid.org/0000-0002-4709-4185};
\\Christopher J. Howland \href{https://orcid.org/0000-0003-3686-9253}{https://orcid.org/0000-0003-3686-9253};
\\Roberto Verzicco \href{https://orcid.org/0000-0002-2690-9998}{https://orcid.org/0000-0002-2690-9998};
\\Detlef Lohse \href{https://orcid.org/0000-0003-4138-2255}{https://orcid.org/0000-0003-4138-2255}.
}

\appendix
\section{Additional experimental details}\label{sec:appa}

\subsection{Fluid density}\label{sec:appa1}

\begin{figure}
  \centering
 \vspace{0cm} 
\includegraphics[width=0.85\columnwidth]{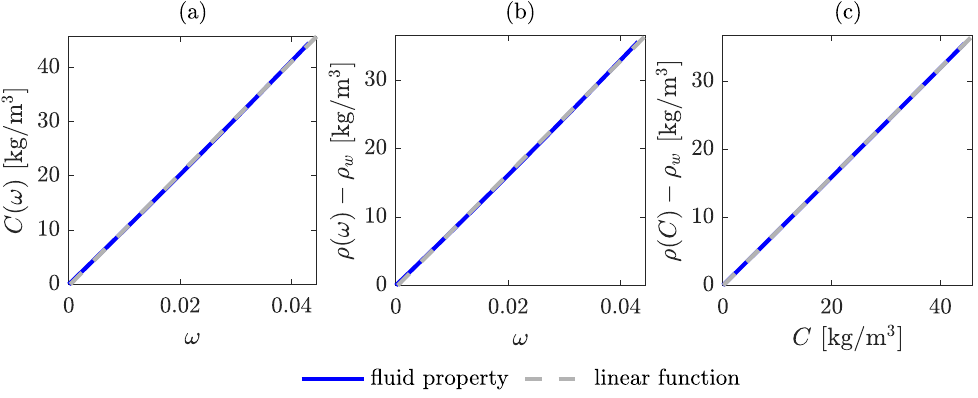}
\caption{\label{fig:fl}
Given the definition~\eqref{eq:defmf} and the empirical correlations~\eqref{eq:novot}-\eqref{eq:nov4}, the relative dependence of density of the solution, $\rho$, mass fraction, $\omega$, and solute concentration, $C$, is obtained (blue solid lines).
We report $C(\omega)$~(panel~a), $\rho(\omega)-\rho_w$~(panel~b) and $\rho(C)-\rho_w$~(panel~c), being $\rho_w$ the water density defined as in~\eqref{eq:refwd}.
The profiles shown here correspond to $\vartheta=25^{\circ} \text{C}$.
The fluid properties are well approximated by linear functions (grey dashed lines).
} 
\end{figure}

Following the empirical correlations of \citet{Novotny1988}, the density of an aqueous solution of KMnO$_4$ can be determined as a function of the solute concentration $C$ (expressed in mol/dm$^3$) and the temperature of the solution $\vartheta$ (expressed in $^{\circ}\text{C}$) as:
\begin{equation}
\rho(C,\vartheta)=\rho_w(\vartheta)+ A_1(\vartheta)C+A_2(\vartheta)C^{3/2}\text{  ,} 
    \label{eq:novot}
\end{equation}
where the water density $\rho_w(\vartheta)$ and the temperature-dependent coefficients $A_1(\vartheta)$ and $A_2(\vartheta)$, are given by:
\begin{eqnarray}
    \rho_w(\vartheta) &=& +9.997\times 10^2\frac{\text{kg}}{\text{m}^3}+2.044\times 10^{-1}\frac{\text{kg}}{^\circ \text{C} \cdot \text{m}^3 }\cdot \vartheta+\nonumber\\
    &&-6.174\times 10^{-2}\frac{\text{kg}}{(^\circ \text{C})^{3/2} \cdot \text{m}^3 }\cdot \vartheta^{3/2}\text{  ,} \label{eq:refwd}
\end{eqnarray}
\begin{eqnarray}    
    A_1(\vartheta) &=& +1.223\times 10^{-1}\frac{\text{kg}}{\text{mol}}-1.029\times 10^{-4}\frac{\text{kg}}{^\circ \text{C}\text{ mol}}\cdot \vartheta+\nonumber\\
    &&+8.093\times 10^{-6}\frac{\text{kg}}{(^\circ \text{C})^2\text{ mol}}\cdot \vartheta^2\text{  ,} 
\end{eqnarray}    
\begin{eqnarray}    
    A_2(\vartheta) &=& -1.485\times 10^{1}\frac{\text{kg}}{\text{m}^3}\left(\frac{\text{dm}^3}{\text{mol}}\right)^{3/2}+9.079\times 10^{-1}\frac{\text{kg}}{^\circ \text{C }\text{m}^3}\left(\frac{\text{dm}^3}{\text{mol}}\right)^{3/2}\cdot \vartheta+\nonumber\\
    &&-7.566\times 10^{-3}\frac{\text{kg}}{(^\circ \text{C }^2)\text{m}^3}\left(\frac{\text{dm}^3}{\text{mol}}\right)^{3/2}\cdot \vartheta^2\text{  .} \label{eq:nov4}
\end{eqnarray}
The relative dependence of density of the solution, $\rho$, mass fraction, $\omega$, and solute concentration, $C$, is shown in figure~\ref{fig:fl} (blue solid lines).
We report here $C(\omega)$~(panel~a), $\rho(\omega)-\rho_w$~(panel~b) and $\rho(C)-\rho_w$~(panel~c), being $\rho_w$ the water density defined as in~\eqref{eq:refwd}.
The density of the mixture (figure~\ref{fig:fl}c), $\rho(C)$, is well approximated by a linear function  of the solute concentration \eqref{eq:rt_eq4a} (dashed grey line).

\begin{figure}
    \centering
        \includegraphics[width=0.95\columnwidth]{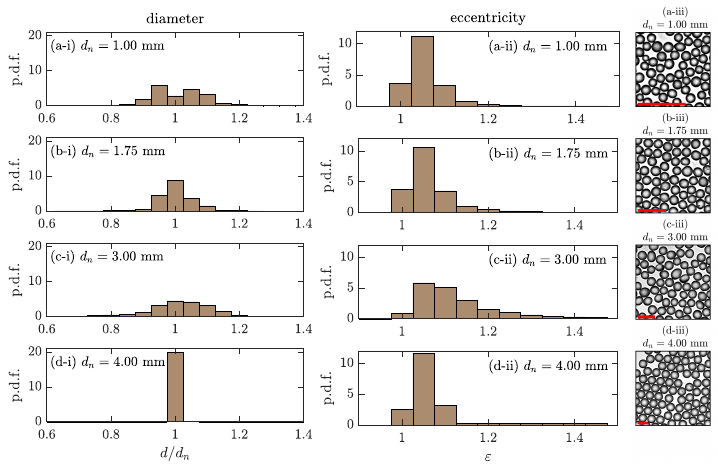}        
    \caption{Beads characterization for all diameters considered. 
    Left column (i): distribution of the diameter of the beads normalized with respect to the nominal diameter ($d_n$). 
    Central column (ii): the eccentricity of the beads ($\varepsilon$) when approximated to ellipses.
    Right column: example of beads employed. A reference length (red bar) corresponding to 5~mm is also indicated.
    }
    \label{fig:bead1}
\end{figure}

\subsection{Characterization of the porous medium}\label{sec:appa0}
The employed glass beads consist of clear, polished glass spheres manufactured by various producers (Hecht Karl, Witeg).
The size distribution of the beads has been determined optically. 
The beads are placed in a transparent container on top of a homogeneous light source (Phlox LEDW-BL $300\times300$~mm$^{2}$).
The images are recorded with a high-resolution camera (Nikon D 850 with lenses Sigma DG Macro 105~mm).
Details are reported in figure~\ref{fig:bead1}.
For all bead size considered, the beads are measured by finding the best-fitting ellipse, and the mean diameter is obtained as $d=\sqrt{ab}$, being $a,b$ the major and minor axis of the ellipse, respectively.
Finally, the shape is also evaluated with the eccentricity, defined as $\varepsilon=a/b$.
A summary of the beads size and shape is reported in figure~\ref{fig:bead1}.
We observe from the statistics (figures~\ref{fig:bead1}c-i,c-ii) and from the pictures (figures~\ref{fig:bead1}c-iii), that the beads having nominal diameter $d_n=3$~mm present a wider distribution of diameters and shapes compared to the other diameters (figures~\ref{fig:bead1}a,b,d) .

\subsection{Correction for the experimental measurement of the mixing length}\label{sec:appa2}
\begin{figure}
    \centering
     \includegraphics[width=0.5\columnwidth]{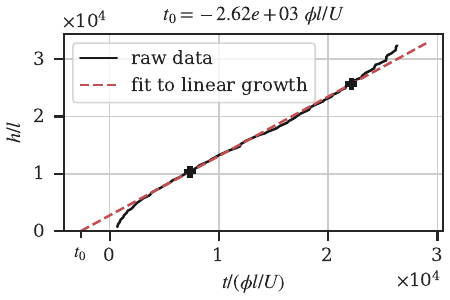}
    \caption{\label{fig:ml2}
    Graphical representation of the correction procedure applied to the experimental measurements.
    Experiments (solid line) are fitted by a linear function (dashed line).
    The new time considered for the experiments is obtained by adding the quantity $t_0$ to the original dimensionless time $t/(\phi\ell/U)$.
    }
\end{figure}

The first instant considered ($t=0$) corresponds to first image analyzed after the experiments has started, i.e. after (i)~the gate has been removed, (ii)~the cell is turned upside-down, and (iii)~the cell is placed in front of the camera. 
This operation produces a delay in the acquisition of the images, and the original time at which the flow starts cannot be accurately determined. 
In addition, the influence of the initial perturbation is apparent.
The initial interface may not be flat, and plumes may form due to the gate opening or due to operational procedure, producing a perturbation of the concentration field.
Since we observe that after an initial transition the evolution of the mixing length is linear, we correct the time used to report the mixing length as follows. 
We first fit the mixing length using a linear function, and then we determine the time $t_0$ at which the value of this function is zero.
The new time considered for the experiments is then obtained by adding the quantity $t_0$ to the original dimensionless time $t/(\phi\ell/U)$.
A graphical representation of this correction process is shown in figure~\ref{fig:ml2}.

\bibliographystyle{jfm}
\bibliography{bibliography}

\end{document}

%% file: table_experiments_dim.tex
E1 && 19.8 & 0.00112 & 0.877 && 1.00 & 0.37 & 7.09$\times 10^{-10}$ && 9.205$\times 10^{-5}$ & 6.632$\times 10^{-6}$ \\ 
E2 && 19.6 & 0.00225 & 1.760 && 1.00 & 0.37 & 7.09$\times 10^{-10}$ && 4.588$\times 10^{-5}$ & 1.331$\times 10^{-5}$ \\ 
E3 && 20.0 & 0.00450 & 3.528 && 1.00 & 0.37 & 7.09$\times 10^{-10}$ && 2.289$\times 10^{-5}$ & 2.667$\times 10^{-5}$ \\ 
E4 && 20.0 & 0.00895 & 7.035 && 1.00 & 0.37 & 7.09$\times 10^{-10}$ && 1.148$\times 10^{-5}$ & 5.319$\times 10^{-5}$ \\ 
\midrule
E5 && 20.0 & 0.00116 & 0.903 && 1.75 & 0.37 & 2.17$\times 10^{-9}$ && 2.921$\times 10^{-5}$ & 2.090$\times 10^{-5}$ \\ 
E6 && 19.6 & 0.00225 & 1.760 && 1.75 & 0.37 & 2.17$\times 10^{-9}$ && 1.498$\times 10^{-5}$ & 4.075$\times 10^{-5}$ \\ 
E7 && 19.7 & 0.00450 & 3.521 && 1.75 & 0.37 & 2.17$\times 10^{-9}$ && 7.489$\times 10^{-6}$ & 8.152$\times 10^{-5}$ \\ 
E8 && 19.5 & 0.00894 & 7.014 && 1.75 & 0.37 & 2.17$\times 10^{-9}$ && 3.759$\times 10^{-6}$ & 1.624$\times 10^{-4}$ \\ 
\midrule
E9 && 20.1 & 0.00111 & 0.868 && 3.00 & 0.35 & 5.07$\times 10^{-9}$ && 1.229$\times 10^{-5}$ & 4.699$\times 10^{-5}$ \\ 
E10 && 19.6 & 0.00235 & 1.837 && 3.00 & 0.35 & 5.07$\times 10^{-9}$ && 5.812$\times 10^{-6}$ & 9.936$\times 10^{-5}$ \\ 
E11 && 20.0 & 0.00453 & 3.547 && 3.00 & 0.35 & 5.07$\times 10^{-9}$ && 3.009$\times 10^{-6}$ & 1.919$\times 10^{-4}$ \\ 
E12 && 19.2 & 0.00905 & 7.098 && 3.00 & 0.35 & 5.07$\times 10^{-9}$ && 1.504$\times 10^{-6}$ & 3.840$\times 10^{-4}$ \\ 
\midrule
E13 && 19.7 & 0.00112 & 0.872 && 4.00 & 0.37 & 1.13$\times 10^{-8}$ && 5.791$\times 10^{-6}$ & 1.054$\times 10^{-4}$ \\ 
E14 && 19.7 & 0.00225 & 1.760 && 4.00 & 0.37 & 1.13$\times 10^{-8}$ && 2.867$\times 10^{-6}$ & 2.129$\times 10^{-4}$ \\ 
E15 && 19.9 & 0.00451 & 3.529 && 4.00 & 0.37 & 1.13$\times 10^{-8}$ && 1.430$\times 10^{-6}$ & 4.269$\times 10^{-4}$ \\ 
E16 && 19.9 & 0.00892 & 7.007 && 4.00 & 0.37 & 1.13$\times 10^{-8}$ && 7.203$\times 10^{-7}$ & 8.475$\times 10^{-4}$ \\ 
\bottomrule

%% file: table_experiments_JFM.tex
E1 & 200 & 0.37 & 558 & 4.535$\times 10^{10}$ & 5.669$\times 10^{3}$ & 2.173$\times 10^{3}$ & 0.289 & 0.0005 \\ 
E2 & 200 & 0.37 & 558 & 9.099$\times 10^{10}$ & 1.137$\times 10^{4}$ & 4.359$\times 10^{3}$ & 0.580 & 0.0010 \\ 
E3 & 200 & 0.37 & 558 & 1.824$\times 10^{11}$ & 2.280$\times 10^{4}$ & 8.737$\times 10^{3}$ & 1.163 & 0.0021 \\ 
E4 & 200 & 0.37 & 558 & 3.637$\times 10^{11}$ & 4.546$\times 10^{4}$ & 1.742$\times 10^{4}$ & 2.320 & 0.0042 \\ 
\midrule
E5 & 114 & 0.37 & 558 & 4.667$\times 10^{10}$ & 3.126$\times 10^{4}$ & 6.846$\times 10^{3}$ & 1.595 & 0.0029 \\ 
E6 & 114 & 0.37 & 558 & 9.099$\times 10^{10}$ & 6.096$\times 10^{4}$ & 1.335$\times 10^{4}$ & 3.110 & 0.0056 \\ 
E7 & 114 & 0.37 & 558 & 1.820$\times 10^{11}$ & 1.219$\times 10^{5}$ & 2.671$\times 10^{4}$ & 6.222 & 0.0112 \\ 
E8 & 114 & 0.37 & 558 & 3.626$\times 10^{11}$ & 2.429$\times 10^{5}$ & 5.320$\times 10^{4}$ & 12.395 & 0.0222 \\ 
\midrule
E9 & 67 & 0.35 & 558 & 4.490$\times 10^{10}$ & 1.515$\times 10^{5}$ & 1.627$\times 10^{4}$ & 5.795 & 0.0104 \\ 
E10 & 67 & 0.35 & 558 & 9.495$\times 10^{10}$ & 3.204$\times 10^{5}$ & 3.441$\times 10^{4}$ & 12.256 & 0.0220 \\ 
E11 & 67 & 0.35 & 558 & 1.834$\times 10^{11}$ & 6.189$\times 10^{5}$ & 6.646$\times 10^{4}$ & 23.672 & 0.0425 \\ 
E12 & 67 & 0.35 & 558 & 3.670$\times 10^{11}$ & 1.239$\times 10^{6}$ & 1.330$\times 10^{5}$ & 47.370 & 0.0850 \\ 
\midrule
E13 & 50 & 0.37 & 558 & 4.506$\times 10^{10}$ & 3.605$\times 10^{5}$ & 3.454$\times 10^{4}$ & 18.393 & 0.0330 \\ 
E14 & 50 & 0.37 & 558 & 9.101$\times 10^{10}$ & 7.281$\times 10^{5}$ & 6.976$\times 10^{4}$ & 37.150 & 0.0666 \\ 
E15 & 50 & 0.37 & 558 & 1.824$\times 10^{11}$ & 1.460$\times 10^{6}$ & 1.398$\times 10^{5}$ & 74.474 & 0.1336 \\ 
E16 & 50 & 0.37 & 558 & 3.622$\times 10^{11}$ & 2.898$\times 10^{6}$ & 2.777$\times 10^{5}$ & 147.861 & 0.2652 \\ 
\bottomrule

%% file: table_simulations_JFM.tex
S1 & 17 & 0.37 & 100 & 5.268$\times 10^{8}$ & 1.000$\times 10^{5}$ & 3.334$\times 10^{3}$ & 5.102 & 0.0510 \\ 
S2 & 17 & 0.37 & 100 & 1.666$\times 10^{9}$ & 3.162$\times 10^{5}$ & 1.054$\times 10^{4}$ & 16.135 & 0.1614 \\ 
S3 & 17 & 0.37 & 100 & 5.268$\times 10^{9}$ & 1.000$\times 10^{6}$ & 3.334$\times 10^{4}$ & 51.024 & 0.5102 \\ 
\midrule
S4 & 35 & 0.37 & 100 & 4.214$\times 10^{9}$ & 1.000$\times 10^{5}$ & 6.669$\times 10^{3}$ & 5.102 & 0.0510 \\ 
S5 & 35 & 0.37 & 100 & 1.333$\times 10^{10}$ & 3.162$\times 10^{5}$ & 2.109$\times 10^{4}$ & 16.135 & 0.1614 \\ 
S6 & 35 & 0.37 & 100 & 4.214$\times 10^{10}$ & 1.000$\times 10^{6}$ & 6.669$\times 10^{4}$ & 51.024 & 0.5102 \\ 
\midrule
S7 & 52 & 0.37 & 100 & 1.422$\times 10^{10}$ & 1.000$\times 10^{5}$ & 1.000$\times 10^{4}$ & 5.102 & 0.0510 \\ 
S8 & 52 & 0.37 & 100 & 4.498$\times 10^{10}$ & 3.162$\times 10^{5}$ & 3.163$\times 10^{4}$ & 16.135 & 0.1614 \\ 
S9 & 52 & 0.37 & 100 & 1.422$\times 10^{11}$ & 1.000$\times 10^{6}$ & 1.000$\times 10^{5}$ & 51.024 & 0.5102 \\ 
\midrule
S10 & 70 & 0.37 & 100 & 3.372$\times 10^{10}$ & 1.000$\times 10^{5}$ & 1.334$\times 10^{4}$ & 5.102 & 0.0510 \\ 
S11 & 70 & 0.37 & 100 & 1.066$\times 10^{11}$ & 3.162$\times 10^{5}$ & 4.218$\times 10^{4}$ & 16.135 & 0.1614 \\ 
S12 & 70 & 0.37 & 100 & 3.372$\times 10^{11}$ & 1.000$\times 10^{6}$ & 1.334$\times 10^{5}$ & 51.024 & 0.5102 \\ 
\midrule